\documentclass[referee]{raa} 
\usepackage{lscape}           
\usepackage{lineno}
\usepackage{natbib}
\usepackage[a4paper, left=25mm, right=25mm, top=20mm, bottom=20mm]{geometry}
\bibstyle{model2-names.bst}
\usepackage{graphicx,times}             
\usepackage{amssymb,amsmath}
\bibpunct{(}{)}{;}{a}{}{,}

\usepackage[a4paper=true,dvipdfm=true,pagebackref=true]{hyperref}
\hypersetup{colorlinks = true, linkcolor = green, anchorcolor = red, citecolor = blue, filecolor = red, pagecolor = red, urlcolor = red}

\begin{document}

   \title{Principal Component Analysis of Ground Level Enhancement of Cosmic Ray
Events
}

 \volnopage{Vol.0 (20xx) No.0, 000--000}      
   \setcounter{page}{1}          

   \author{R.E. Ugwoke
      \inst{1,2*}
      \and A. A. Ubachukwu
      \inst{2}
      \and J. O. Urama
      \inst{2}
      \and O. Okike
      \inst{3}
      \and J. A. Alhassan
      \inst{2}
      \and A. E. Chukwude
       \inst{2}
   }

   \institute{Department of Physics, Federal University of Technology Owerri, Imo state Nigeria\\
        \and
             Department of Physics and Astronomy, University of Nigeria, Nsukka, Nigeria\\
             \and
                 Department of Industrial Physics, Ebonyi State University, Abakaliki, Nigeria\\
{\it *romanusejike1971@gmail.com and romanus.ugwoke@futo.edu.ng}\\ 
\vs\no
   {\small Received~~20xx month day; accepted~~20xx~~month day}}

\abstract{We applied principal component analysis (PCA) to the study of five ground level enhancement (GLE) of cosmic ray (CR) events. The nature of the multivariate data involved makes PCA a useful tool for this study. A subroutine program written and implemented in R software environment generated interesting principal components. Analysis of the results shows that the method can distinguish between neutron monitors (NMs) that observed Forbush decrease (FD) from those that observed GLE at the same time. The PCA equally assigned NMs with identical signal counts with the same correlation factor ($r$) and those with close r values equally have a close resemblance in their CR counts. The results further indicate that while NMs that have the same time of peak may not have the same $r$, most NMs that had the same r also had the same time of peak. Analyzing the second principal components yielded information on the differences between NMs having opposite but the same or close values of $r$. NMs that had the same $r$ equally had the tendency of being in close latitude.
}

   \authorrunning{R.E. Ugwoke \& A.A. Ubachukwu \& J.O. Urama \& O. Okike \& J. A. Alhassan \& A.E. Chukwude} 
   \titlerunning{Principal component analysis of ground level enhancement of cosmic ray events}  

   \maketitle

\linenumbers
\renewcommand\makeLineNumber{}
%
\section{Introduction}           
\label{sect:intro}

Cosmic ray (CR) studies have been carried out by a lot of researchers who have sought for the understanding of their  origins, acceleration mechanisms, time profiles, and their interactions in the heliosphere, magnetosphere as well as atmosphere. Their effects both on the geomagnetic storms and lightning have provided a lot of insights that enable accurate predictions of weather and weather forecasts necessary for air transports, and for space explorations using satellites and cosmonauts.
The time profile of CRs which either indicates a sudden (or sharp) drop in their intensity on arrival or  a rapid increase in their intensity is always studied along with other geomagnetic indices.
There have been cases of a concurrent sudden increase in intensity with time called ground level enhancement (GLE) in some neutron monitor stations along with a sharp decrease in intensity with time also known as Forbush decrease (FD) in other stations for a given event.

Researchers have employed different methods in their analysis of both FD and GLE but we do not know of any application of principal component analysis (PCA) in the study of GLE. PCA has however been used extensively in other fields of sciences and engineering, as a tool for investigations.

The technique of reducing the dimension of a multivariate data set such that the variance of the data set is maximized is called PCA \citep[e.g.][]{FG:2004,OC:2011,TI:2015,JC:2016,Ka:2017}. 
The hidden structures in the entire data are the principal components (PCs) \citep{Ha:2018} and all the PCs are orthogonal \citep[e.g.][]{OC:2011,J:2002}.  The PCs retain almost all the information in the original data set \citep{SD:2008,OC:2011,JC:2016}. The first PC (PC1) has the highest variance \citep[e.g.][]{OC:2011,FJ:2019} and therefore contains most of the information in the original data set.

PCA is done either by the singular value decomposition method or by the method of Covariance Matrix \citep{Mo:1981,MN:2006,DR:2008,SD:2008,TI:2015}. The first two are sufficient for interpretation if their percentage variance is up to 80\% \citep{OC:2011}.

Identification of Cosmic TeV Gamma-ray protons during extensive air showers was carried out by \citet{FG:2004} by the method of PCA. This they did by applying the PCA on the two-dimensional particle density fluctuations, which according to their report provided a decreasing sequence of covariance matrix eigenvalue with features that are not the same for different primary CRs.
\citet{MN:2006}, improved on the combination of PCA with visualization method. They were able to control all steps of the visualization pipeline using PCA results and thus removing the limitation of making the combination a preprocessing step of dimension reduction.

\citet{OC:2011} carried out a multivariate study of FD simultaneity and were able to use PCA to discriminate between globally simultaneous and non-simultaneous FDs. Their work suggests that the structure of the  intensity time profile of FDs measured at different locations appear to be similar if there are strong positive correlations between raw data and PC1.
Ear recognition using blocked-based PCA along with Decision Fusion by \citet{TI:2015} achieved faster recognition performance compared to previous methods.

\citet{ZC:2019} also applied PCA to their assessment of the impacts of meteorological elements on the concentration of particulate matter (PM10). They were able to distinguish three PCs and how they affect PM10 in the four seasons of the year.
\citet{RG:2020} applied PCA methods to the CR data of a newly installed muon detector to remove the atmospheric temperature effect on the data. The PC regression they applied captured at least 77\% of the variability in the data due to the effect of the temperature of the air layers at the site of the detector.

\citet{OA:2022} investigated the relations between automated selected FD, worldwide lightning frequency, sunspot number and other solar-terrestrial drivers using PCA. They were able to show the level of correlation of FD with these parameters. The authors also compared the result of their FD catalogue with that of Pushkov Institute of Terrestrial Magnetism, Ionosphere, and Radio Wave Propagation, Russian Academy of Sciences (IZMIRAN) with PCA. They were able to identify similarities and differences between the FDs correlations in the two catalogues with the said solar/geophysical parameters.

\begin{figure}[htbp]
\begin{subfigure}{1.0\textwidth}
	\centerline{\includegraphics[width = 0.8\textwidth]{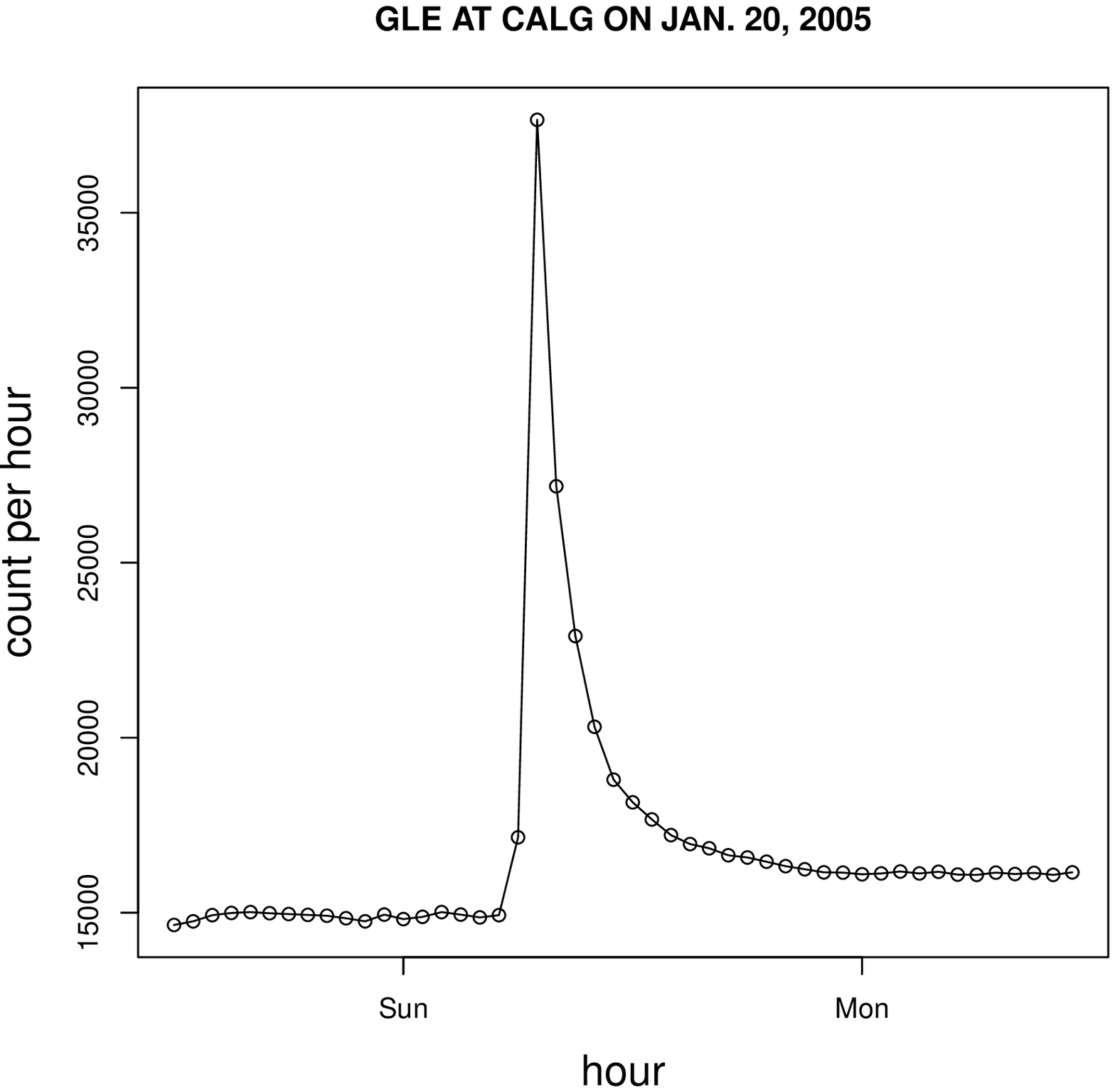}}
	\caption{GLE event of January 20, 2005 at CALG.}
	\label{Figure 1a}
	\end{subfigure}
	
\begin{subfigure}{1.0\textwidth}
	\centerline{\includegraphics[width = 0.8\textwidth]{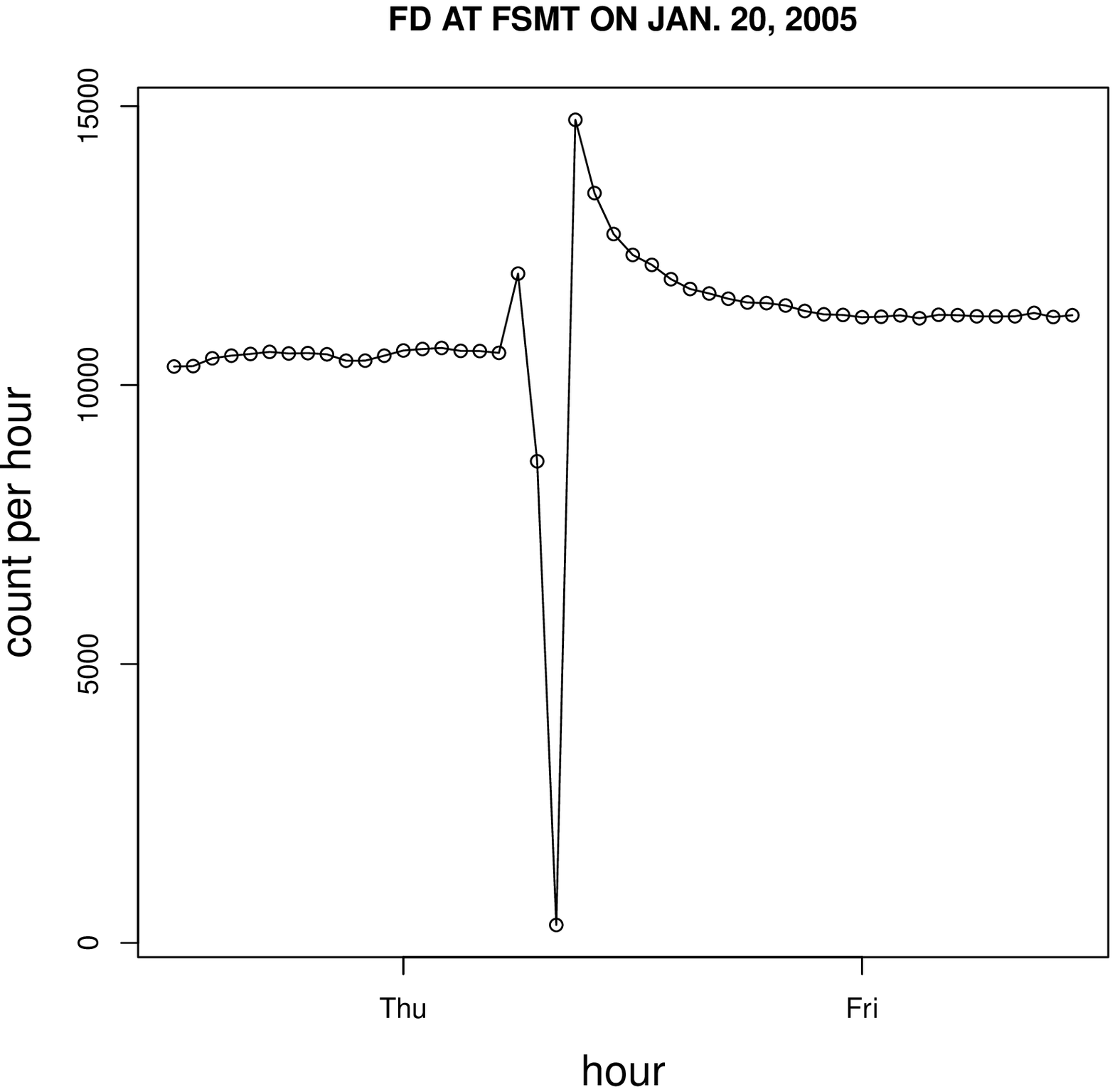}}
	\caption{FD observed by FSMT during the GLE of January 20, 2005.}
	\label{Figure 1b}
	\end{subfigure}
	
\end{figure}

In the present submission, we present a principal component analysis of the GLEs of May 7, 1978; September 29, 1989; October 22, 1989; April 15, 2001 and January 20, 2005. These events are also known as GLEs 31, 42, 44, 60 and 69 respectively in \citet{Mms:2012}. Given the numerous applications of PCA to scientific investigations as have  been cited above, we strongly believe that the application of the method to the analysis of GLE multivariate data will not only yield interesting results but open doors to further research on GLE. Figure \ref{Figure 1a} shows the time profile of a typical GLE event that occurred on January 20, 2005, and was observed at Calgary (CALG) NM. The Figure shows that the galactic cosmic ray (GCR) count before  the GLE was stable and that the GLE commenced with a small increase in CR count and thereafter rose to peak count. The decay was equally quasi-exponential. Figure \ref{Figure 1b} is another example; this time FD occurred the same day as the GLE and observed by NM at FortSmith (FSMT).

\subsection{THEORY OF PRINCIPAL COMPONENT ANALYSIS}\label{Theory}
The under-listed equations by \citet{JC:2016} succinctly describe the theory of PCA applied in this investigation.
For an n x p data matrix X, with n-dimensional vectors $x_{1}$, $x_{2}$, ..., $x_{p}$, here there are p numerical variables. The jth column is the vector $x_{j}$ of observations on the jth variable.\\
Let $X_{a}$ be a linear combination of the columns of the matrix having maximum variance such that 
\begin{equation} \label{eqn}
Xa = \sum_{j=1} ^{p} a_{1j}x_{j}
\end{equation}
$a$ is a vector of constants $a_{1}$, $a_{2}$, ..., $a_{p}$.
The variance is given by 
\begin{equation} \label{eqn}
var(Xa) = {a}^{\prime}Sa
\end{equation}
where $^{\prime}$ is the transpose and $S$ is the covariance matrix associated with $Xa$. \citet{J:2002} and \citet{JC:2016} recommended the normalization required for the solution to be\\
\begin{equation} \label{eqn}	
	{a}^{\prime}{a} = 1 
\end{equation} 
This ensures that\\
\begin{equation} \label{eqn}
var(Xa) = {a}^{\prime}Sa = \lambda{a}^{\prime}a = \lambda
\end{equation} 
where $a$ is the eigenvector and $\lambda$ is the eigenvalue. 
For linear combinations of uncorrelated $Xa_{k}$, having maximum variance,\\
 \begin{equation} \label{eqn}
Xa_{k} =  \sum_{j=1} ^{p} a_{jk}x_{j} 
\end{equation} 
This means that as long as $k^{\prime} \neq k$,
\begin{equation} \label{eqn}
a^{\prime}_{k}Sa_{k} = \lambda_{k}a^{\prime}_{k}a_{k} = 0
\end{equation}
$Xa_{k}$ is the PCs, while the elements of the eigenvector $a_{k}$ are the PC loadings. In this work, the PC loadings are the correlation of the PC with the raw data. The elements of the linear combination $Xa_{k}$ are the PC scores which in this work is the time series signal of the CR counts. 

It has been a common practice to use mean-centred variables ($x^{\star}_{j}$ ) in PCA   \citep{J:2002,OC:2011,JC:2016} where\\
\begin{equation} \label{eqn}
x^{\star}_{j} = x_{ij}-\overline{x}_{j}
\end{equation}
and $\overline{x}_{j}$ is the mean value of the variable $j$.
To avoid the problems associated with the units of all the $p$ variables which may not be the same, \cite{JC:2016} recommended that in addition to mean centering of the variables, $x_{ij}$ is divided by the standard  deviation of the  n observations of variable $j$. Thus 
\begin{equation} \label{eqn}
Z_{ij} = \frac{x_{ij}-\overline{x}_{j}}{s_{j}}
\end{equation}
where $s_{j}$ is the standard deviation. The matrix $X$ is therefore replaced with $Z_{ij}$ and $Xa_{k}$ replaced with $Za_{k}$ so that
\begin{equation} \label{eqn}
Za_{k} =  \sum_{j=1} ^{p} a_{jk}z_{j}
\end{equation} 
This last equation results in PCs that do not change for any linear transformation of the units. In this case too, the normalization used is
\begin{equation} \label{eqn}
	\tilde{a}^{\prime}_{k}\tilde{a}_{k} = \lambda_{K}, K = 1, 2, ..., p.\\
	\end{equation}
 and not that in equation (3). With this normalization, the coefficient of correlation ($r$) between $k^{th}$ PC and the $j^{th}$ variable is given as\\
\begin{equation} \label{eqn}
r_{var_{j}},PC_{k} = \sqrt{\lambda_{k}a_{jk}}
\end{equation} 
Thus with the normalization $\tilde{a}^{\prime}_{k}\tilde{a}_{k} = \lambda_{K}$ instead of ${a}^{\prime}{a} = 1$, The coefficient of the new loading $\tilde{a}_{k}$ are the correlation between each original variable and the $k^{th}$ PC. 

\section{DATA AND METHOD OF DATA ANALYSIS.}
In the World-Wide NM Network \url{http://cro.izmiran.ru/common/links.htm} there are catalogues of raw CR data covering decades of CR observations. The website which is rich in pressure-corrected hourly and minute data is hosted by Pushkov Institute of Terrestrial Magnetism, Ionosphere, and Radio Wave Propagation, Russian Academy of Sciences (IZMIRAN). We selected from this website composite hourly data from NM stations spread across all latitudes.

Nineteen and twenty-three NMs that had composite data were studied in the May 7, 1978, and October 22, 1989 events respectively. In the September 29, 1989 event, we selected twenty-seven NM stations. There were twenty-eight NMs selected for the April 15, 2001 event while in the January 20, 2005 event, twenty-two NM stations were selected for the study. In all the events under study, forty-eight hours of observation (twelve hours before, twenty-four hours on the day of the event and twelve hours after the day of the event) were considered.
 
We employed awk programming to convert the data frame from the above website to a data frame readable in an R software environment. R is a software used in data science mainly for statistical analysis. Many packages are embedded in it and could be obtained from \url{https://cran.R-project.org}. \citet{VS:2022} described R as an integrated suite of software facilities for data manipulation, calculation and graphical display. 

Identifying the hidden pattern or information in a multivariate data set with correlated variables is the main reason for PCA of such data \citep[e.g.][]{J:2002,OC:2011,Ka:2017}. Dimensionality reduction of the data set to variables of lower dimensions (PCs) through the following steps as shown in \citet{Skp:2020} is the procedures involved:

\begin{itemize}
\item centering the data set, 
\item Computing the covariance and correlation matrix, 
\item Computation of the eigenvectors and the eigenvalues,
\item Computing the PC scores and loadings.
\end{itemize}

These PCs are orthogonal and carry nearly all the information in the original data set with PC1 having the maximum variance followed by PC2 up to the last one \citep[e.g.][]{Jo:2002,OC:2011,Ka:2017}. The interpretation is based on the correlation of the original data set with the signal, see \citet{OC:2011} since variables with common correlation values share some common information. A subroutine program we wrote and implemented in the R software environment generated the PC loadings and scores shown in Figures 1 to 10 and used them for the interpretations.

\section{RESULTS AND DISCUSSION}
\subsection{INTERPRETATION OF THE CORRELATION OF THE PCs WITH THE RAW DATA}
The PCA results are presented in Figures \ref{Fig. 2}-\ref{Fig. 11}. The upper and lower panels of each of these diagrams stand for the correlation coefficients of the PCs with the raw data and the variation patterns of the PCs respectively. The y-axes of the upper and lower panels are, thus, labeled ``correlation" and ``Signal variations". The physical interpretation/representation of the signal variations indicated in the lower panels depends on the values of the correlation coefficient ($r$) in the upper panels of each of the diagrams. Strong signals will register high values of r (statistically significant) whereas the weak signals will be associated with small or non-significant $r$. With regard to CR intensity variation which is the subject of the current work, PCs associated with large values of $r$ might be global GLEs whereas those with smaller values of $r$ are representative of local or station dependent GLEs. These global GLEs are seen at the same time by all the NMs irrespective of their locations. The signal variation plotted on the lower panel is the form or the average variation pattern of CR raw data observed by the NM stations at various locations of the Earth. \cite{oh:08} and \cite{OC:2011} would refer to such events as globally simultaneous CR variations. The PCA theory and some details of the mathematical relationship between the correlation of the PCs and the raw data are presented in Section \ref{Theory}. 

As suggested by \cite{OC:2011}, another useful statistics, besides the correlation coefficient, used to decided the simultaneity of an event is the percentage variance. The first principal component (PC1) is usually the one with the highest variance and as such, contains most of the information in the original data \citep[]{OC:2011,Ka:2017,OA:2022}. Any event whose PC1 is associated with about 80\% variance would be taken as a strong and simultaneous GLE. For easy reference, the values of r associated with each PC1 of the five events analyzed here are presented in Tables \ref{table:1}, \ref{table:3}, \ref{table:5}, \ref{table:7} and \ref{table:9} whereas the variances associated with all the PCs for each of the events are presented in Tables \ref{table:2}, \ref{table:4}, \ref{table:6}, \ref{table:8}, and \ref{table:10}.

\subsection{CORRELATION OF THE FIRST PRINCIPAL COMPONENT (PC1) WITH RAW DATA ON MAY 7, 1978 GLE (GLE 31).}

\begin{figure}[htbp]
	\centerline{\includegraphics[width = 0.8\textwidth]{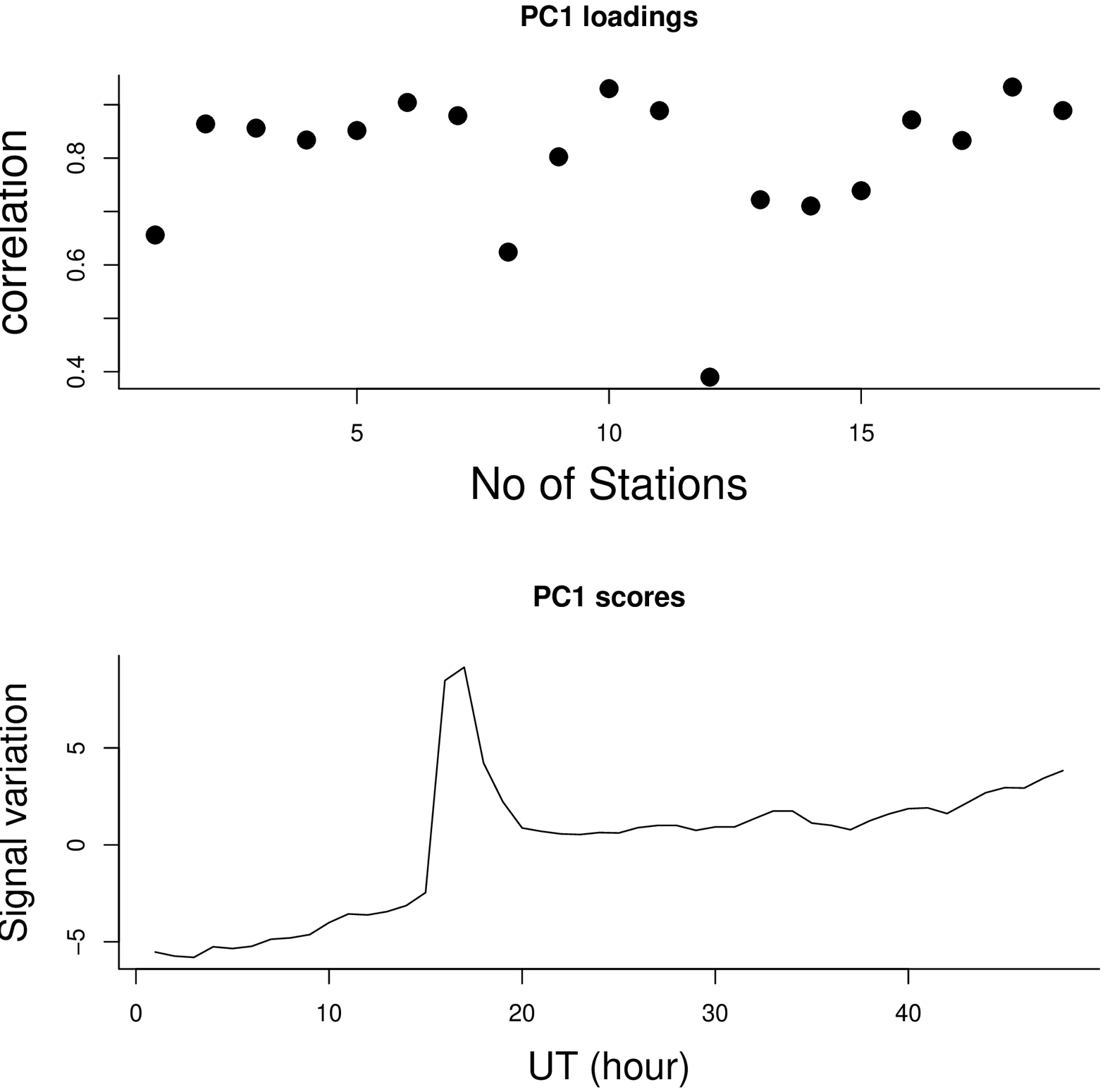}}
	\caption{PC1 for GLE of May 07, 1978 event.}
	\label{Fig. 2}
\end{figure}

The PC1 for the event of 7/05/1978 is presented in Fig. \ref{Fig. 2} and the quantitative results are reported in Tables \ref{table:1} and \ref{table:2}. The total variance associated with the PC1 is 65.52. This implies that PC1 accounts for 65.52\% intensity variation in the raw data at the time of this event. This value of variance associated with PC1 is less than the 80\% benchmark of \citet{OC:2011}. The event may not, therefore, be regarded as globally simultaneous. The variance reported in Table \ref{table:2} shows that about the first three PCs (65.52 + 19.58 + 0.07998 = 85.18\%) are required to account for the CR intensity variation for this event at all the stations. The contributions of each of the remaining 16 PCs (1.48\%, 1.20\%, 1.0\%, 0.74\%, 0.61\%, 0.53\%, 0.3\%, 0.3\%, 0.2\%, 0.2\%, 0\%) are far less and may safely be regarded as noise. The general trend as seen from the signal in PC1 of this event is that the majority of the NMs had a gradual rise and fall in CR counts (in other words small FDs) before the main day of the event and never had an instant rise to peak level count. Many of them had pre-peak counts that are not evenly spaced. Their decay also never reversed to the original background GCR counts. 

Individual plots of CALG, HRMS, INVK, and SOPO (plots not shown) which had r values of 0.86, 0.83, 0.85 and 0.83 respectively show that they all had enhancements in CR count before the peak. Their peaks occurred at 4:00 UT, 3:00 UT, 4:00 UT and 4:00 UT respectively. No instant jump to the peak was observed in the GLE event in these stations. The initial stage of the event however varied in them. All of them also show that the enhancement occurred after some series of FD. Their decay was not all that quasi exponential and they could not decay down to the initial GCR count before the GLE.

IRKT, and YKTK had a common r value (0.90) and the same time of peak (4:00 UT). Both of them had quite odd CR counting characterized by rising and falling CR counts up to the peak and also during their decay. It is not only that they could not decay to the original GCR count but the rising and falling CR count continued after the decay. Their profile is much more similar than with any other NM. However, TXBY and MGDN with r = 0.92 had the closest profile to the two of them.

SNAE and MTWS had the same value of $r$ (0.89) and peak time (4:00 UT). Their profiles are also quite similar. In each of them, there was evidence of an FD before the GLE and they both had pre-peak counts. Whereas they had a quasi-exponential decay, they did not decay down to their previous GCR count before the event. GSBY with $r$ = 0.85 had an initial gradual rise in CR count after which it jumped to the peak count at 4:00 UT. Its decay was gradual and not necessarily quasi-exponential. After the decay, it established a new level of GCR count that is much higher than what it was before the GLE. 

In this event too, JUN1 had $r$ = 0.88. It had a more rapid rise to peak count than the NMs with r = 0.89 and equally a faster decay than MTWS. While SNAE and MTWS had a pre-peak count, JUN1 did not. JUN1 had peak at 03:00 UT.

PTFM and ROME having $r$ = 0.69 and 0.72 respectively, are almost the same in their GLE CR count. They also had multiple small FDs before the little enhancement. Their enhancement was quite minimal compared to the rest so far considered. In two of them, the count in CRs decayed instantly to a value close to the last part of the recovery phase of the precursory FD and started rising and falling again in the same pattern before the GLE. LMKS and OULU with $r$ = 0.80 and 0.70 respectively also had instant rise to peak count. The two NMs recorded small FD before the GLE. The decay in LMKS was more rapid than others in this event. After about two decay counts, it came down to the GCR level. On the other hand, the decay in OULU  was also a two steps down to the GCR count but its first decay was too close to the peak count.

APTY and KIEL  which had instant rise to peak count also had close $r$ values equal to about 0.60 and 0.65 respectively. In them, the GCR count was relatively flat (without an FD) before the GLE event. The decay to GCR count which was almost instantaneous occurred the same way in them.

The count in NVBK was unexpected. Though it had $r$ = 0.40 and should have been very much like APTY, it was a complete FD before a small enhancement. Being a deviation from the rest, it should have a negative correlation like those in GLE 69 that had FD when others were having GLE. Perhaps there may be an error in the data entry of the archive because the count before the peak in NVBK is a four-digit figure as against all others in the station with five-digit figures. It is also seen that except for JUN1, all the NMs that had r less than 0.85 had a peak at 03:00 UT but the peak of NVBK occurred at 04:00 UT. 
 
\begin{table}
	\caption{Full names of the place where the NMs for the May 7, 1978 event (GLE 31) are located, their short names and the $r$ values of the PC1}
	\label{table:1}
	\centering
	\begin{tabular}{rlll}
		\hline
	SN. & NM full name & $NM short name$ & $r$ \\ 
		\hline
	    1 & APATITY & APTY & 0.65\\
        2 & CALGARY & CALG & 0.86\\
        3 & GOOSEBAY & GSBY & 0.85\\
        4 & HERMANUS & HRMS & 0.83 \\
	    5 & INUVIK & INVK & 0.85\\
        6 & IRKUTSK & IRKT & 0.90\\
        7 & JUNGFRAUJOCH & JUN1 & 0.88\\
        8 & KIEL & KIEL & 0.60\\
	    9 & LOMNICKYSTIT & LMKS & 0.80\\
       10 & MAGADAN & MGDN & 0.92\\
       11 & MTWASHINGTON & MTWS & 0.89\\
       12 & NOVOSIBIRSK & NVBK & 0.40\\
       13 & OULU & OULU & 0.70\\
       14 & POTCHEFSTROOM & PTFM & 0.69\\
       15 & ROME & ROME & 0.72\\
       16 & SANAE & SNAE & 0.89\\
       17 & SOUTHPOLE & SOPO & 0.83\\
       18 & TIXIEBAY & TXBY & 0.92\\
       19 & YAKUTSK & YKTK & 0.90\\

	\hline
\end{tabular}
\end{table}

\subsection{CORRELATION OF THE SECOND PRINCIPAL COMPONENT (PC2) WITH RAW DATA ON MAY 7, 1978 GLE (GLE 31)}

\begin{figure}[htbp]
	\centerline{\includegraphics[width = 0.8\textwidth]{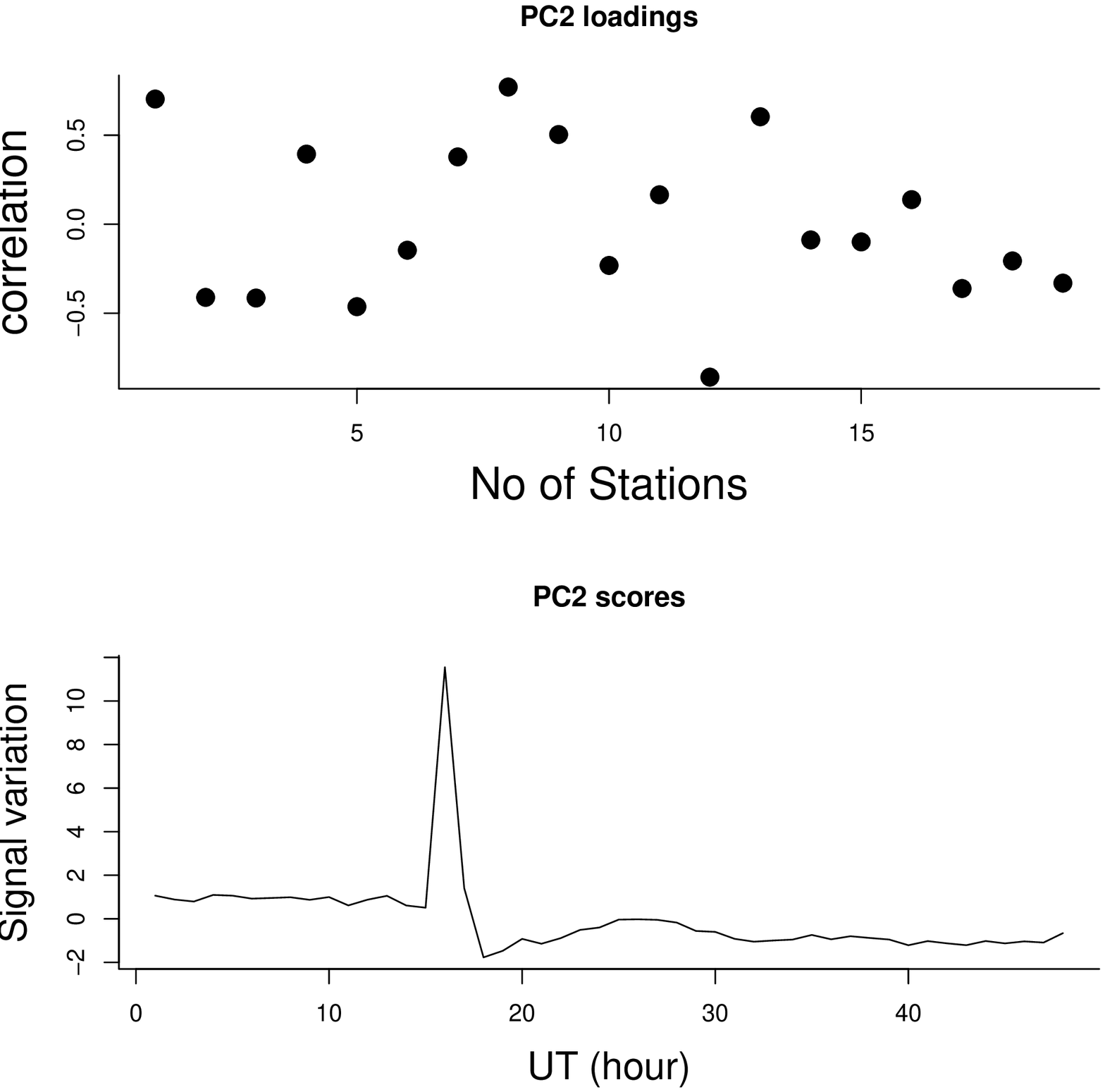}}
	\caption{PC2 for GLE of May 07, 1978 event.}
	\label{Fig. 3} 
\end{figure}

The PC2 (Figure \ref{Fig. 3}) of May 7, 1978 accounts for 19.58\% of the information in the raw data (see Table \ref{table:2}). The relevant information needed in this PC is those for whom their $r$ values have counterparts that their $r$ sign are opposite and are equal or close to being equal.

APTY, KIEL and OULU all had $r$ = 0.55, 0.55 and 0.51 respectively. In PC1 they equally shared close values of $r$. All of them had an instant jump to a high peak count. Their decay equally took the same pattern. NVBK on the other hand had $r$ = -0.58 and an FD at the time others had enhancement in CR counts. Though it had positive $r$ in PC1, it was the least $r$ because of this departure from the trend seen in others.

CALG, GSBY and INVK in this PC2 had $r$ = -0.45 while SOPO and YKTK had $r$ = -0.40. When these two groups are compared with those having$r$ = 0.40 (JUN1) and 0.50 (LMKS), it is seen that each of these groups already had close values of $r$ in PC1. The group with positive $r$ had instant jump to peak count and almost dropped to the GCR count after two decay counts. On the other hand, those with negative $r$ values did not instantaneously jump to peak count. In these latter groups of NMs, the CR count also took several decay steps down to the GCR counts.

PTFM and ROME had $r$ = -0.2 and also had similar CR counts at their rising and decay phases. Both of them differed from SNAE and MTWS which had $r$ = 0.20 in having instant decay to almost the GCR count. SNAE and MTWS had a gradual decay count.

HRMS which had $r$ = 0.40 recorded an instant rise to peak count and almost decayed to the GCR count instantly, while MGDN with $r$ = -0.30 had several intermediate counts before the peak and also decayed gradually to GCR counts. In this PC2, having an opposite sign that are equal or almost equal suggest that there exist a feature in them that are dissimilar.

\begin{table}
	\caption{PCA summary of the May 07, 1978 (GLE 31) event}
	\label{table:2}
	\centering
	\begin{tabular}{rllll}
		\hline
	SN. & Principal Component & Standard deviation & Proportion of Variance & Cumulative Proportion\\ 
		\hline

1 & PC1 & 3.5282       & 0.6552    &  0.6552      \\
2 & PC2 & 1.9285       & 0.1958    &  0.8509     \\
3 & PC3 & 1.23277      & 0.07998   &  0.93090     \\
4 & PC4 & 0.5304       & 0.0148    &  0.9457      \\
5 & PC5 & 0.47765      & 0.01201   &  0.95771     \\
6 & PC6 & 0.43608      & 0.01001   &  0.96772      \\                                  
7 & PC7 & 0.37367      & 0.00735   &  0.97507     \\         
8 & PC8 & 0.34156      & 0.00614   &  0.98121     \\                                                      
9 & PC9 & 0.31574      & 0.00525   &  0.98646      \\                  
10 & PC10 & 0.23667    & 0.00295   &  0.98940      \\                   
11 & PC11 & 0.2267     & 0.0027    &  0.9921      \\
12 & PC12 & 0.20279    & 0.00216   &  0.99427     \\                                   
13 & PC13 & 0.1950     & 0.0020    &  0.9963      \\                    
14 & PC14 & 0.14501    & 0.00111   &  0.99738     \\                                   
15 & PC15 & 0.13299    & 0.00093   &  0.99831     \\                 
16 & PC16 & 0.11915    & 0.00075   &  0.99906      \\                                    
17 & PC17 & 0.10477    & 0.00058   &  0.99964     \\                                  
18 & PC18 & 0.07219    & 0.00027   &  0.99991      \\                                    
19 & PC19 & 0.04099    & 0.00009   &  1.00000     \\
\hline
\end{tabular}
\end{table}

\subsection{CORRELATION OF THE FIRST PRINCIPAL COMPONENT (PC1) WITH RAW DATA IN THE SEPTEMBER 29, 1989 EVENT (GLE 42)}

\begin{figure}[htbp]
	\centerline{\includegraphics[width = 0.8\textwidth]{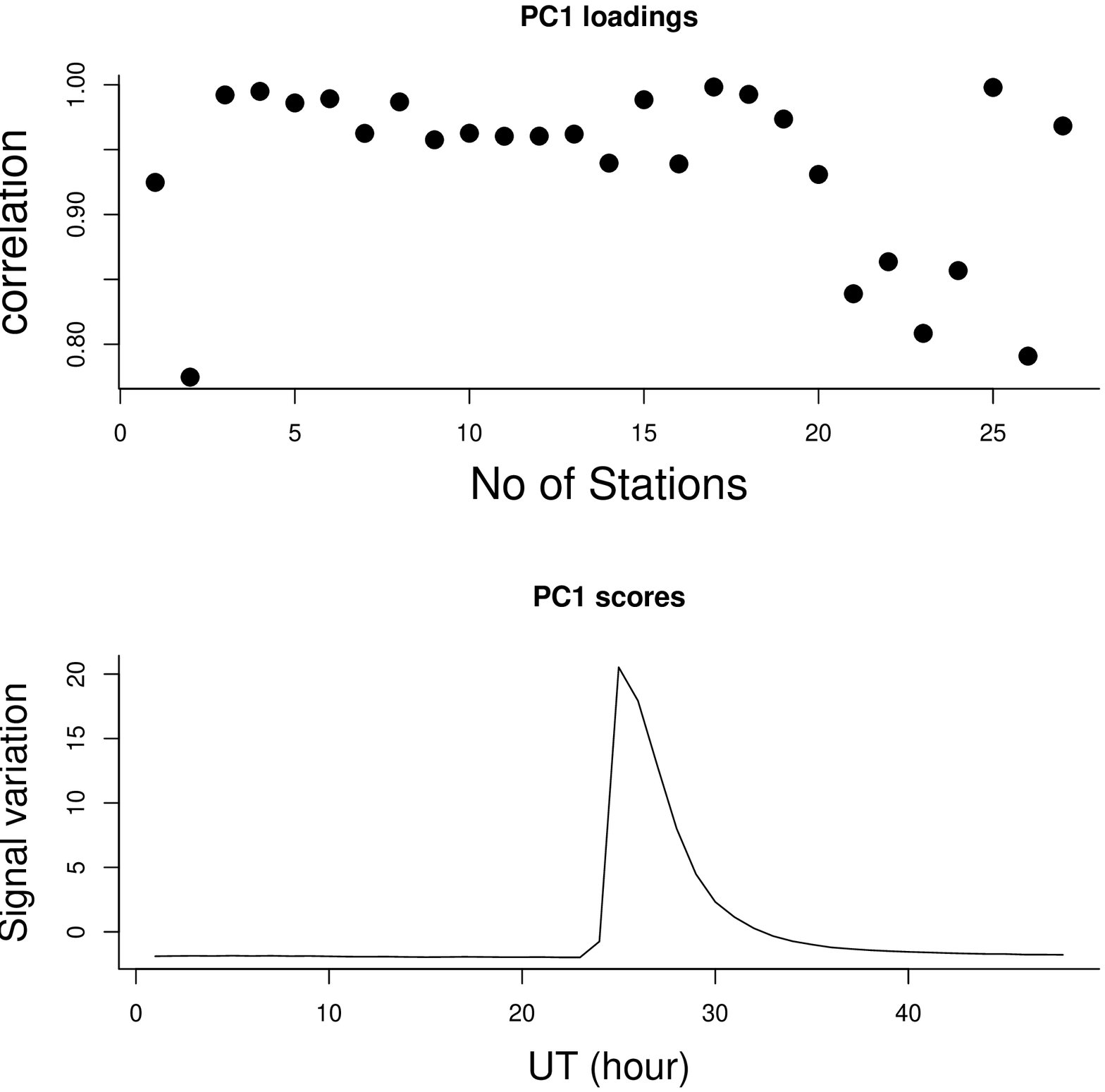}}
	\caption{PC1 for ground level enhancement of September 29, 1989 event.}
	\label{Fig. 4}
\end{figure}
Figure \ref{Fig. 4} represents the PC1 of the September 29, 1989 GLE event. The results of the  GLE event are quantitatively shown in Tables \ref{table:3} and \ref{table:4}. It is interesting to note that the correlation coefficient, $r$, of each station with the PC1 is very high ($>0.7$, Table \ref{table:3}). Table \ref{table:4} equally shows that the variance associated with PC1 is very high (0.8825), higher than the 80\% benchmark of \cite{OC:2011}. This is an indication that the event is very strong and globally observed by all the NMs no matter where they are located on Earth. The strong nature of this event as suggested by PC1 is in perfect agreement with the submission of \citet{Mc:2014}. 

The PC loading is shown as a signal representing how each of the signal from the individual NMs would appear. The PC1 in this event assigned the same $r$ to NMs having the same signal pattern. In general, the decay is quasi-exponential, though some tended to be more closely related than others.
  
Table \ref{table:3} shows the correlation of the first principal component PC1 with raw data. The values are approximated values derived from the PC loading and score in Figure \ref{Fig. 3}. In this table, all the NMs had high correlation coefficient suggesting that their individual signal for the CR count are almost the same with the signal in Figure \ref{Fig. 3}. 

Individual plots of the NMs (plots not shown here) show that for those with $r$ values from 0.97 to 1.0, their CR counts both the GCR and the GLE counts up to the peak are quite similar with little differences. THUL, CAPS and MGDN had $r$ = 1.0 and are all in the high latitude Northern hemisphere. They equally had the same profile and peaked at 12:00 UT. They differed from others that had $r$ from 0.99 in having the first decay count that is too close to the peak. Similar to this group with $r$ = 1.0, are CALG, KIEL and NVBK (three of which had $r$ = 0.99). This group too had a pre-peak count that are too close to the peak value and are in the upper mid latitude in the Northern hemisphere.

The $r$ values of GSBY, INVK, TXBY and NWRK are 0.98, 0.98, 0.97 and 0.97 respectively. Only GSBY had peak at 12:00 UT while the other three had theirs at 13:00 UT. All of them had similar CR counting structure in their initial, main and decay phases. They are more like those that had $r$ = 0.99 except that their initial increase in CR count is greater compared with those that had $r$ = 0.99. DPRV (with $r$ = 0.8) is just like this group except that its pre-peak count is closer to the peak just like those that had $r$ = 0.99. Only INVK and TXBY are in the high latitude in the Northern hemisphere while the rest (including DPRV) are in the mid latitude Northern hemisphere. Except for INVK, the other three are equally at close longitude.

IRKT, IRK2, IRK3, JUN1, HRMS, JUN1 and JUNG had $r$ = 0.95. They had similar profile; a relatively small enhancement in the first one hour of the GLE, followed by the peak (at 12:00 UT) and quasi-exponential decay. All the NMs in this category are in the mid latitude Northern hemisphere except HRMS that is in the Southern hemisphere.

MCMD and KERG had $r$ = 0.93. Their CR count varied slightly. Their initial count was very small similar to those with $r$ = 1.0. While MCMD had two other counts before the peak, KERG had one more count before the peak which is widely separated from the peak count. Their peak did not occur the same time and while KERG is at mid latitude Southern hemisphere, MCMD is at high latitude Southern hemisphere.

OULU and APTY are at high latitude Northern hemisphere. They are also at close longitude and had their peak at 13:00 UT. They have two peaks or two counts at their peaks beside a widely separated pre-peak count. Two of them had $r$ = 0.92. SNAE (with $r$ = 0.86) and TERA (with $r$ 0.85) have identical profile. Their GLE began with very small increase in CR count and later two other steps in their counts before the peak that occurred at 14:00 UT. Both of them are at the high latitude in the Southern hemisphere. 

PTFM, TBLS, TSMB and BJNG had 0.84, 0.80, 0.78 and 0.73 as their respective $r$ values. They had similar profile characterised by a rapid rise to peak count and rapid decay too. While TBLS and BJNG are at lower part of the mid latitude in the Northern hemisphere, PTFM and TSMB are at the low latitude in the Southern hemisphere.

From the forgoing it is seen that in PC1 there are differences in the rising pattern of some NMs with close $r$ values. Looking at their decay pattern, these differences tend to disappear. The decay count of all the NMs having $r$ = 0.97 to 1.00 are quite similar and there are about 60\% of them. All the entire NMs having high $r$ values arise from their decay pattern being the same. Their different rises to peak count are also responsible for their various $r$ values.

In GLE 42, thirteen of the NMs that had peak at 12:00 UT did not have $r$ values that are the same. In addition eight of the NMs that had peaks at 13:00 UT did not have similar $r$ values. The same trend is observed in the NMs that had peak at 14:00 UT. All these not withstanding, NMs with the same $r$ values tend to be simultaneous.  This is in agreement with  \citep{OC:2011} who were able to distinguish between  NMs with simultaneous Forbush decrease with those that are not simultaneous.

\begin{table}
	\caption{Full names of the place where the NMs for September 29, 1989 event (GLE42) are located, their short names and their correlation coefficient ($r$ values) of the PC1}
	\label{table:3}
	\centering
	\begin{tabular}{rlll}
		\hline
	SN. & NM full name & $NM short name$ & $r$ \\ 
		\hline
	1 & APATITY & APTY & 0.92\\
	2 & BEIJING & BJNG & 0.73\\
	3 & CALGARY & CALG & 0.99\\
	4 & CAPESSHMIDT & CAPS & 1.00\\
	5 & DEEPRIVER & DPRV & 0.98\\
	6 & GOOSEBAY & GSBY & 0.98\\
	7 & HERMANUS & HRMS & 0.95 \\
	8 & INUVIK & INVK & 0.98\\
	9 & IRKUTSK2 & IRK2 & 0.95\\
	10 & IRKUTSK3 & IRK3 & 0.95\\
	11 & IRKUTSK & IRKT & 0.95\\
	12 & JUNGFRAUJOCH & JUN1 & 0.95\\
	13 & JUNGFRAUJOCH2 & JUNG & 0.95\\
	14 & KERGUELEN & KERG & 0.93\\
	15 & KIEL & KIEL & 0.99\\
	16 & MCMURDO & MCMD & 0.93\\
	17 & MAGADAN & MGDN & 1.00\\
	18 & NOVOSIBIRSK & NVBK & 0.99\\
	19 & NEWARK & NWRK & 0.97\\
	20 & OULU & OULU & 0.92\\
	21 & POTCHEFSTROOM & PTFM & 0.84\\
	22 & SANAE & SNAE & 0.86\\
	23 & TBILISI & TBLS & 0.80\\
	24 & TERREADELIE & TERA & 0.85\\
	25 & THULE & THUL & 1.00\\
	26 & TSUMEB & TSMB & 0.78\\
	27 & TIXIEBAY & TXBY & 0.97\\
	
	\hline
\end{tabular}
\end{table}

\subsection{CORRELATION OF THE SECOND PRINCIPAL COMPONENT (PC2) WITH RAW DATA IN GLE 42.}

\begin{figure}[htbp]
	\centerline{\includegraphics[width = 0.8\textwidth]{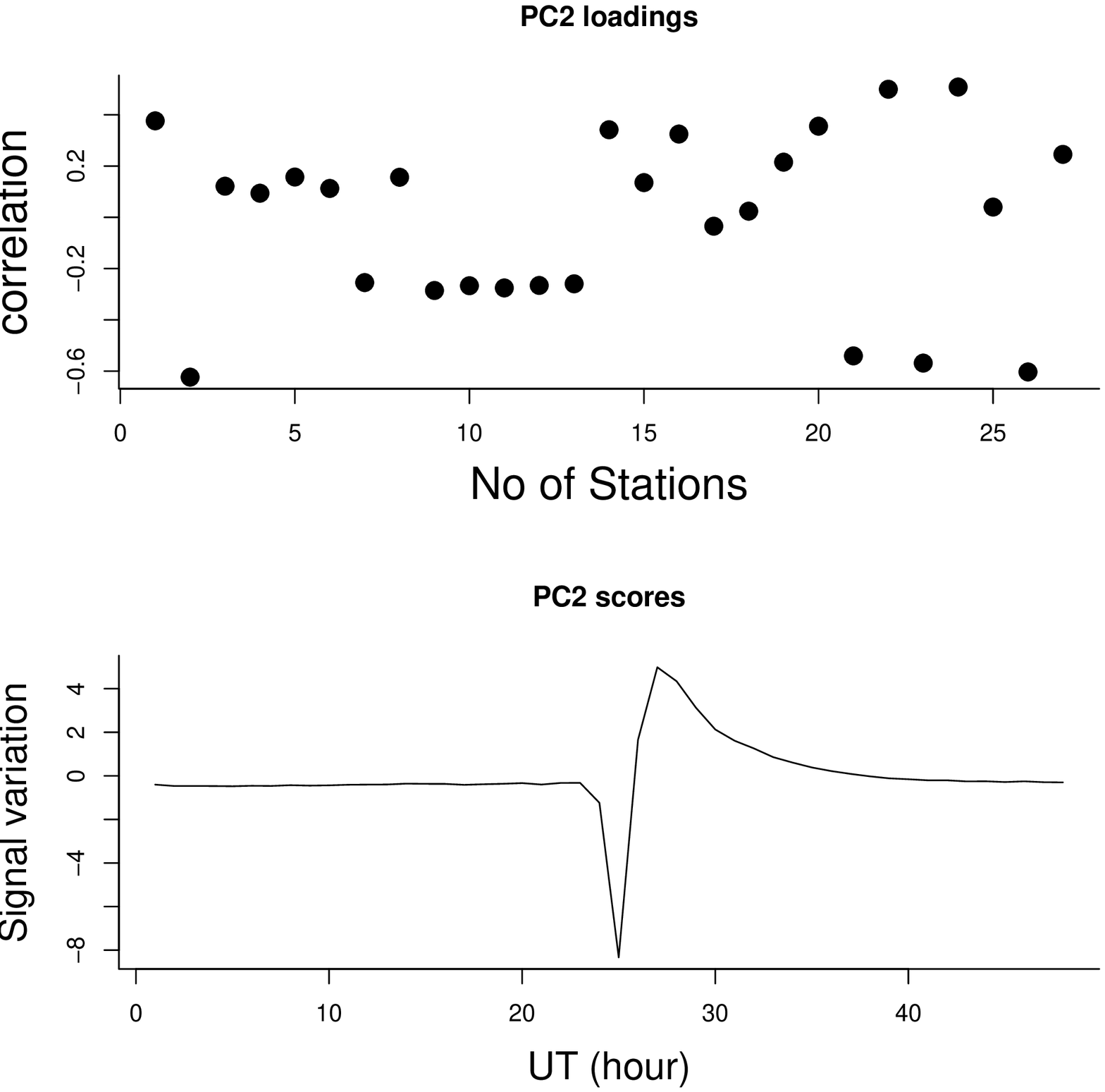}}
	\caption{PC2 for GLE of September 29, 1989 event.}
	\label{Fig. 5}
\end{figure}

PC2 in this event (shown in Figure \ref{Fig. 5}) accounts for 11.16\% of the total intensity variation in the raw data as can be seen from its total variance (11.16\%) in Table \ref{table:4}. In PC2 of this event, the features of the NMs CR counts prior to the GLE and the main phase are reflected by their $r$ values. The main phase here include the moment of initial increase in CR counts up to the moment decay commenced. 

IRKT, IRK2, IRK3, JUN1, JUNG, HRMS and TBLS with $r$ = -0.3 showed no sign of double peak or pre-peak count. In hourly plot, double peaks are not obvious.
KERG and MCMD having $r$ = 0.3, showed features of seemingly double peaks with little time lag between them.

For the following NMs; DPRV, NVBK, TXBY and KIEL whose $r$ are 0.11, 0.00, 0.20 and 0.10, their rising pattern of CR counts are similar with greater time lag between the pre-peak and peak count (or possibly between the two peaks) when compared with those having $r$ = 0.3. OULU with $r$ = 0.4 had a larger time lag between the pre-peak and peak count than those with $r$ = 0.3.

TSMB, TBLS and BJNG all had $r$ = -0.6. Each of them had an initial rise in CR count and rose from there to their peak count. In this case, they did not have pre-peak count. In comparison, those whose {\color{red}$r$} is 0.4 (APTY, OULU, SNAE, TERA), had pre-peak count or possibly double peaks. 

\begin{table}
	\caption{PCA summary of the September 29, 1989 event (GLE42)}
	\label{table:4}
	\centering
	\begin{tabular}{rllll}
		\hline
	SN. & Principal Component & Standard deviation & Proportion of Variance & Cumulative Proportion\\ 
		\hline

1 & PC1 &  4.8814 & 0.8825 & 0.8825\\
2 & PC2 &  1.7358 & 0.1116 & 0.9941\\
3 & PC3 &  0.32775 & 0.00398 & 0.99811\\
4 & PC4 &  0.15356 & 0.00087 & 0.99898\\
5 & PC5 &  0.12043 & 0.00054 & 0.99952\\
6 & PC6 &  0.06744 & 0.00017 & 0.99969\\
7 & PC7 &  0.06494 & 0.00016 & 0.99984\\
8 & PC8 &  0.03861 & 0.00006 & 0.99990\\                  
9 & PC9 &  0.02789 & 0.00003 & 0.99993\\
10 & PC10 & 0.02225 & 0.00002 & 0.99995\\
11 & PC11 & 0.01999 & 0.00001 & 0.99996\\                  
12 & PC12 & 0.01713 & 0.00001 & 0.99997\\
13 & PC13 & 0.01222 & 0.00001 & 0.99998\\
14 & PC14 & 0.01181 & 0.00001 & 0.99998\\
15 & PC15 & 0.009556 & 0.000000 & 0.999990\\
16 & PC16 & 0.009313 & 0.000000 & 0.999990\\
17 & PC17 & 0.008321 & 0.000000 & 0.999990\\
18 & PC18 & 0.006835 & 0.000000 & 0.999990\\
19 & PC19 & 0.00602 & 0.000000 & 1.000000\\
20 & PC20 & 0.005639 & 0.000000 & 1.000000\\
21 & PC21 & 0.005134 & 0.000000 & 1.000000\\
22 & PC22 & 0.004496 & 0.000000 & 1.000000\\
23 & PC23 & 0.004206 & 0.000000 & 1.000000\\
24 & PC24 & 0.003346 & 0.000000 & 1.000000\\              
25 & PC25 & 0.003096 & 0.000000 & 1.000000\\
26 & PC26 & 0.002781 & 0.000000 & 1.000000\\
27 & PC27 & 0.002326 & 0.000000 & 1.000000\\
  \hline
\end{tabular}
\end{table}

\subsection{CORRELATION OF THE FIRST PRINCIPAL COMPONENT (PC1) WITH RAW DATA IN OCTOBER 22 1989 EVENT (GLE 44)}

\begin{figure}[htbp]
	\centerline{\includegraphics[width = 0.8\textwidth]{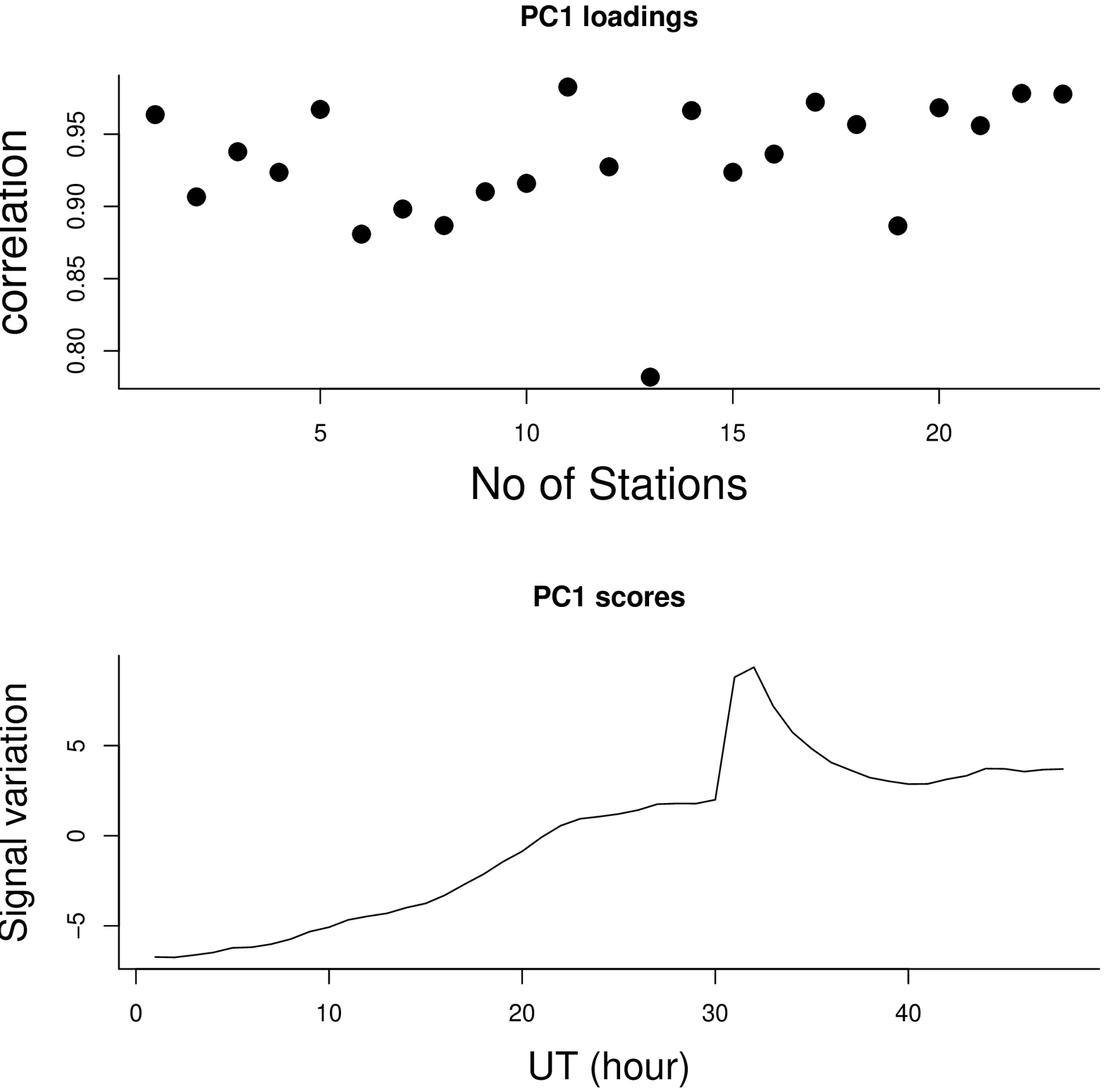}}
	\caption{PC1 for GLE of October 22, 1989 event.}
	\label{Fig. 6}
\end{figure}
The PC1 of this event is as shown in Figure \ref{Fig. 6} while Tables \ref{table:5} and \ref{table:6} contains the quantitative results. The total variance associated with the PC1 alone is 86.81. This also imply that 86.81\% of the total intensity variation in the raw data in all the stations for this event is accounted for by PC1. Since this value exceeds the benchmark of \citet{OC:2011}, the event may be regarded as a globally simultaneous event. The contributions of the other PCs, especially PC3 to PC23 (1.74\%, 1.35\%, 0.62\%, 0.19\%, 0.13\%, 0.09\%, 0.06\%, and the rest are 0.00\% ) suggest that they are noise.

The PC1 signal of the October 22, 1989 event (see Figure \ref{Fig. 6}) represents the combined signals from all the NMs. Generally the signal shows that there was a major FD on Saturday 21 of October and its recovery continued until later part of Sunday when the GLE started. There were other small FDs as part of the recovery phase of the large FD. It is also seen from the signal that there was a pre peak count.  Usually, this signal is a representation of the manifestation of signals from majority of the NMs, thus not all of them will have the pre peak count.

APTY, KERG, MGDN, OULU, TXBY, INVK, and YKTK, TXBY had $r$ = 0.97 suggesting that their signal structure of the CR count are similar. All of them except MGDN had their peak at 19.00 UT while MGDN had its peak at 18:00 UT. Before the peak count, there was another count in all of them which is an enhancement of high value except in MGDN whose initial count was just a recovery from a minor FD prior to the enhancement. The CR count before the peak in INVK and TXBY are widely separated more than as are in the rest in this group. Five of these NMs had similar decay which cannot be said to be quasi-exponential.  MGDN is the only one that is at mid latitude in the Southern hemisphere while the rest are in the high latitude in the Northern hemisphere. OULU whose $r$ value is 0.98 has the same profile as APTY, INVK, KERG and TXBY NMs and is also in the high latitude in the Northern hemisphere. TERA that had r = 0.96 is very much like those with $r$ = 0.97 both in its rise to peak count and decay count. It however differed from them in the two minor FDs that are part of the recovery phase of the main FD. TERA is in the high latitude in the Southern hemisphere.

NWRK, SNAE and THUL all of which had $r$ = 0.95 have counts that are almost the same as the group above  with $r$ = 0.97. The main difference between them is that in this group, the count before the peak (pre peak count) and the peak count are much closer in time than in the former group. They too did not have at least a quasi-exponential decay. In this group, only NWRK is at lower part of the mid latitude in the Northern hemisphere while the rest are at the high latitude in Northern hemisphere, with exception of SNAE that is in the Southern hemisphere. 

LMKS, CLMX,  HRMS, and NVBK had counts that could not be said to be a GLE. Their $r$ values are 0.93, 0.93, 0.92, and 0.93 respectively and their time of peak are 19:00 UT, 18:00 UT, 20:00 UT and 19:00 UT. Though the recovery phase of the prior large FD was completed in them, the additional rise in the CR count was a gradual increase up to the peak. The decay in each of them was also a rising and falling CR counts. CLMX  looked more like those that had $r$ = 0.95, however both its peak and the first decay count were too close. LMKS and CLMX are at the lower part of the mid latitude in the Northern hemisphere. While HRMS is at the mid latitude in the Southern hemisphere, NVBK is at the upper part of the mid latitude in the Northern hemisphere.

The signal of the CR count in IRKT,  IRK2 and IRK3 are identical though their respective $r$ are 0.89, 0.87 and 0.90 respectively. The three of them apparently had what appears to be double peaks with their corresponding time of peaks as shown; IRKT (11:00 UT, 18:00 UT),  IRK2 (10:00 UT, 18:00 UT) and IRK3 (10:00 UT, 18:00 UT). Their rising count could not show a full recovery from the large FD. These shared parameters could be because their latitudes and longitudes are almost the same. JUN1 and JUNG with $r$ = 0.91 and 0.92 are among those that had what appears to be double peaks like those whose $r$ value are between 0.89 - 0.90. Their slight difference is more in their decay.

CALG, MCMD and SOPO with $r$ values equal to 0.91, 0.75 and 0.89 respectively all had instant jump to high peak count. Though they all showed the large FD prior to the GLE, the minor FDs are barely noticeable. Their decay were equally quasi-exponential. The three of them had simultaneous peak at 18:00 UT. In this event too, we see that NMs with the same $r$ values tend to be simultaneous and 80\% of NMs that had $r$ greater or equal to 0.93 had peak at the same time (19:00 UT).   

\begin{table}
	\caption{Full names of the place where the neutron monitors for the October 22, 1989 event (GLE 44) are located, their short names and their correlation coefficient ($r$ values) of the PC1}
	\label{table:5}
	\centering
	\begin{tabular}{rlll}
		\hline
	SN. & NM full name & $NM short name$ & $r$ \\ 
		\hline
	1 & APATITY & APTY & 0.97\\
	2 & CALGARY & CALG & 0.91\\
	3 & CLIMAX & CLMX & 0.93\\
	4 & HERMANUS & HRMS & 0.92\\
	5 & INUVIK & INVK & 0.97\\
	6 & IRKUTSK2 & IRK2 & 0.87\\
	7 & IRKUTSK3 & IRK3 & 0.90\\
	8 & IRKUTSK & IRKT & 0.89\\
	9 & JUNGFRAUJOCH & JUN1 & 0.91\\
	10 & JUNGFRAUJOCH2 & JUNG & 0.92\\
	11 & KERGUELEN & KERG & 0.97\\
	12 & LOMNICKYSTIT & LMKS & 0.93\\
	13 & MCMURDO & MCMD & 0.75\\
	14 & MAGADAN & MGDN & 0.97\\
	15 & NOVOSIBIRSK & NVBK & 0.93\\
	16 & NEWARK & NWRK & 0.94\\
	17 & OULU & OULU & 0.97\\
	18 & SANAE & SNAE & 0.95\\
	19 & SOUTHPOLE & SOPO & 0.89\\
	20 & TERREADELIE & TERA & 0.96\\
	21 & THULE & THUL & 0.95\\
	22 & TIXIEBAY & TXBY & 0.97\\
	23 & YAKUTSK & YKTK & 0.97\\

	\hline
\end{tabular}
\end{table}

\subsection{CORRELATION OF THE SECOND PRINCIPAL COMPONENT (PC2) WITH RAW DATA IN OCTOBER 22 1989 EVENT (GLE 44)}

\begin{figure}[htbp]
	\centerline{\includegraphics[width = 0.8\textwidth]{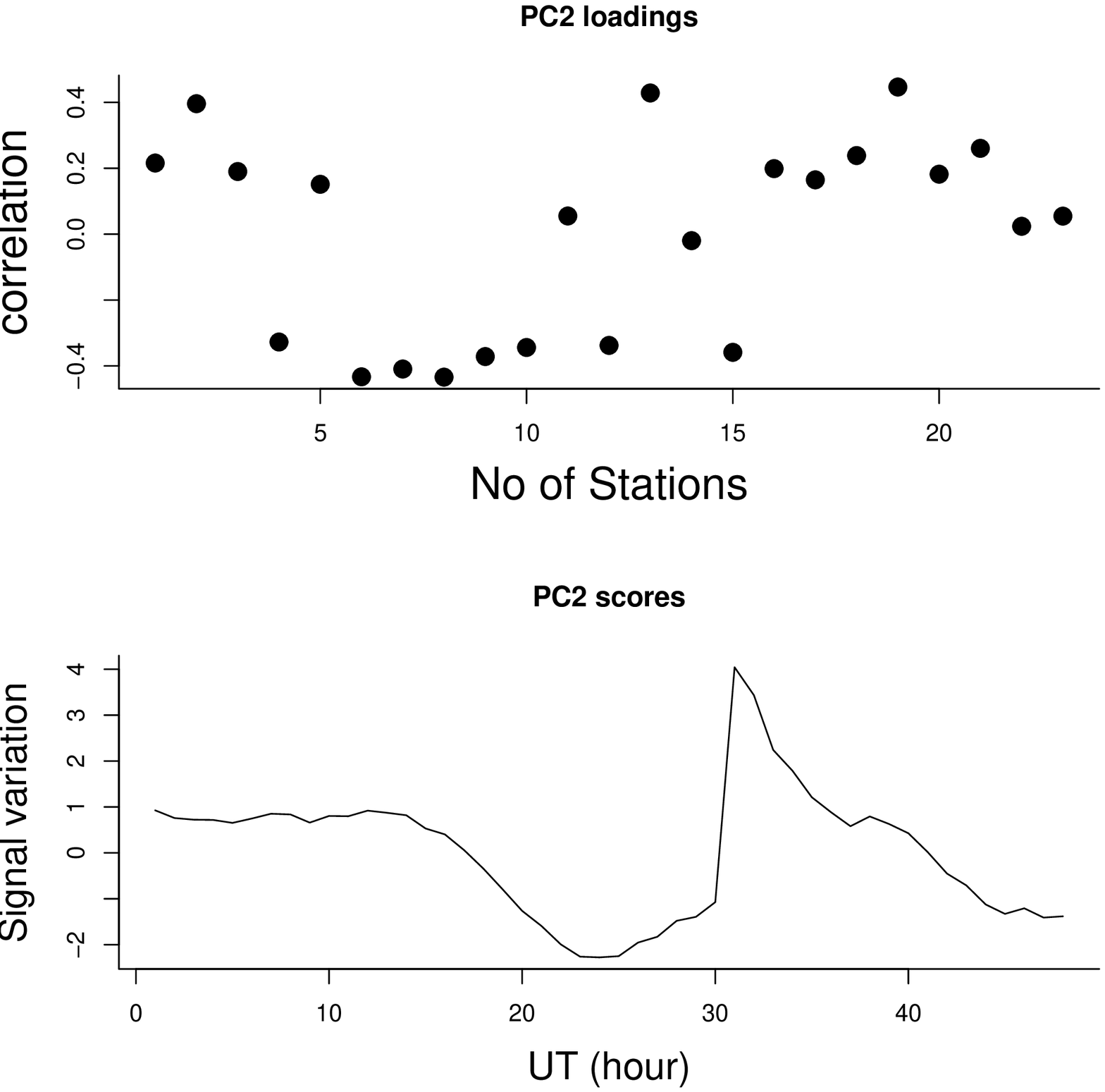}}
	\caption{PC2 for GLE of October 22, 1989 event.}
	\label{Fig. 7}
\end{figure}

8.81\% of the total intensity variation in the raw data is accounted for by PC2 (Figure \ref{Fig. 7}) in this event. In this PC2, CALG, MCMD, and SOPO have $r$ = 0.40, 0.43 and 0.43 respectively. They all had instant rise to peak count. On the other hand, IRK2, IRK3, and IRKT also had $r$ = -0.41, -0.40 and -0.41 respectively. They are the exact opposite of of the the group above having $r$ = 0.4 in that they could not be said to have GLE as they had gradual rise from the previous FD and could not recover to the initial GCR count before the decay commenced.

Similar to this group having negative correlation factor, NVBK, LMKS and HRMS all of which had $r$ = -0.35 had gradual rise in the CR counts up to the maximum count in each of them. They too did not decay to the initial GCR count. JUN1 and JUNG whose $r$ values are -0.39 and -0.36 share the same signal structure as other NMs with negative $r$ above.

APTY, CLMX, NWRK, OULU and TERA had $r$ = 0.2 and all of them had clear GLE unlike those with negative $r$ values. In addition, they had a relatively large initial increase in CR count before the peak. SNAE that has $r$ = 0.25 shared similar profile as NMs with $r$ = 0.2. In general, PC2 in this event show that positive $r$ were assigned to NMs that had clear GLE while negative $r$ are for those that did not have clear GLE.

\begin{table}
	\caption{PCA summary of the October 22, 1989 event}
	\label{table:6}
	\centering
	\begin{tabular}{rllll}
		\hline
	SN. & Principal Component & Standard deviation & Proportion of Variance & Cumulative Proportion\\ 
		\hline

1 & PC1 & 4.4683  & 0.8681 & 0.8681\\
2 & PC2 & 1.42355  & 0.08811 & 0.95619\\
3 & PC3 & 0.63173  & 0.01735 & 0.97354\\
4 & PC4 & 0.55692  & 0.01349 & 0.98702\\
5 & PC5 & 0.37825  & 0.00622 & 0.99324\\
6 & PC6 & 0.2090  & 0.0019 & 0.9951\\
7 & PC7 & 0.17545  & 0.00134 & 0.99648\\
8 & PC8 & 0.1439  & 0.0009 & 0.9974\\                  
9 & PC9 & 0.11568  & 0.00058 & 0.99796\\
10 & PC10 & 0.10044 & 0.00044 & 0.99840\\
11 & PC11 & 0.08689 & 0.00033 & 0.99873\\                  
12 & PC12 & 0.08279 & 0.00030 & 0.99903\\
13 & PC13 & 0.06835 & 0.00020 & 0.99923\\
14 & PC14 & 0.06400 & 0.00018 & 0.99941\\
15 & PC15 & 0.06078 &  0.00016 & 0.99957\\
16 & PC16 & 0.05662 &  0.00014 & 0.99971\\
17 & PC17 & 0.04354 &  0.00008 & 0.99979\\
18 & PC18 & 0.03480 &  0.00005 & 0.99984\\
19 & PC19 & 0.03351 &  0.00005 & 0.99989\\
20 & PC20 & 0.02885 &  0.00004 & 0.99993\\
21 & PC21 & 0.02638 &  0.00003 & 0.99996\\
22 & PC22 & 0.02399 &  0.00003 & 0.99998\\
23 & PC23 & 0.01901 &  0.00002 & 1.00000\\

  \hline
\end{tabular}
\end{table}

\subsection{CORRELATION OF THE FIRST PRINCIPAL COMPONENT (PC1) WITH RAW DATA IN APRIL 15, 2001 (GLE 60)}
\begin{figure}[htbp]
	\centerline{\includegraphics[width = 0.8\textwidth]{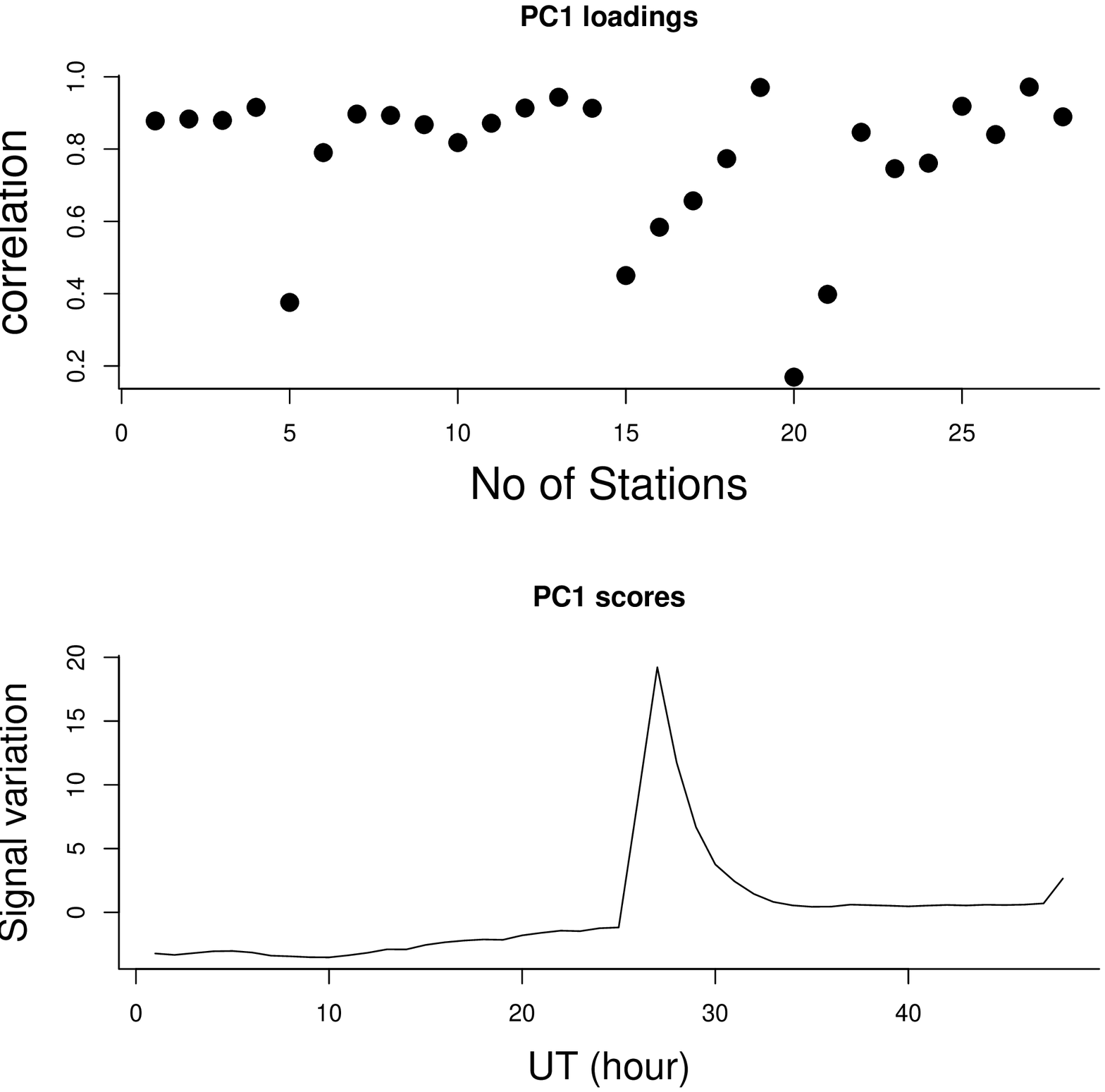}}
	\caption{PC1 for GLE of April 15,2001 event.}
	\label{Fig. 8}
\end{figure}
Figure \ref{Fig. 8} is the PC1 of GLE 60. The quantitative results of this event are in Tables \ref{table:7} and \ref{table:8}. The respective variance associated with PC1, PC2 and PC3 are 64.67, 14.26 and 11.04. This shows that three of these PCs are required to account for 89.97\% (64.67\% + 14.26\% + 11.04\%) of the total intensity variation in the raw data. This is an indication that this is a weak GLE and are seen at different times at different NM stations.

 The PC1 signal shows the combined effect of all the NMs CR count. The correlation in PC1 shows the relationship between the signal from each NM with the raw data. The higher the correlation value of each NM with the raw data signal, the closer the signal of the NM resembles the raw signal (PC1 signal). 

In this event, NVBK had the least $r$ (0.18). It observed several small FDs before the GLE which made it significantly different from any other NM in this group. It rose to instant peak count at 14:00 UT and had a quasi-exponential decay phase.

FSMT and NWRK  whose $r$ values are equal ($r$=0.40) have identical signal structure of the CR hourly count. In their individual plots that are not shown here, they revealed a small FD prior to the GLE. The plots are characterized by instant rise to peak count and quasi-exponential decay. FSMT showed peak at 16:00 UT while NWRK had peak at 14:00 UT. MCMD with the closest $r$ value to FSMT and NWRK had a sharp rise to peak count which occurred at 17:00 UT. It is not known why it should have its peak latter than all the other NMs. Its onset began with a rise to a relatively high count in the CR before it dropped to a value which could be called an FD. From this maximum depression of the FD, it rose instantly to the peak count. 

MWSN which also had instant rise to peak count  at 14:00 UT and $r$ value of 0.65 showed a small FD that is barely noticeable before the GLE. It decayed quasi-exponentially. The $r$ value of MGDN is 0.60. It recorded similar CR count signal as MWSN but the small FD in MGDN is more outstanding than that in MWSN. MGDN also had peak at 14:00 UT.

Both SOPO and NAIN had $r$ = 0.78 and thus similar signal structure of the CR count. The plots for the two stations show a rise to peak count at 14:00 UT, followed by quasi-exponential decay. The precursory FDs that were seen in other NMs did not show up in them.

HRMS and JUN1 also show similar profile characterized by rising and falling CR count during the recovery phase of a precursory large FD, a jump to peak count that occurred at 14:00 UT, and decay phase that is not exponential. The profile from both NMs revealed another small FD beside the large FD before the GLE. The $r$ for both stations is 0.78. 

The $r$ value of APTY, CAPS, and KERG is 0.88. All the profiles show pre-peak counts and a peak count that occurred at 15:00 UT. The background GCR was relatively flat in all of them implying that they did not observe the precursory FD. The decay was quasi-exponential in the three NMs. With the $r$ value of THUL, OULU, CALG and YKTK equal to 0.89, their profile are almost the same with those  characterized by $r$ = 0.88. Apart from YKTK that had an initial small rise in CR count, others had instant rise to peak count that occurred at 14:00 UT. Their profiles did not show any FD just before the GLE and the decay was quasi-exponential in all the profiles.

There are two groups that had $r$ = 0.90 and both groups showed peaks at 14:00 UT. The first group which include INVK, KGSN, and TERA did not show any FD and rose to peak count instantaneously. Their decay was equally quasi-exponential. The second group include LMKS, IRKT and CLMX. In them, there was a large FD that began a day before the GLE and a small FD just before the GLE. They rose to instant peak count at 14:00 UT from the recovery phase of the small FD. They too had quasi-exponential decay phase. 

NRLK and TXBY had the highest $r$ value of 1.00. They both had two small FDs before the GLE and also pre-peak counts. Their peak occurred at 15:00 UT and they decayed quasi-exponentially. In nearly all the NMs in this event, equal $r$ value resulted in simultaneous peak count and similar signal structure of the CR count. Apart from the rising phase of the GLE CR count, the decay phase of FSMT, MCMD, HRMS, JUNE1 IRKT and LMKS are the reason for which their $r$ values are low. This is because the decay show a significant departure from the trend displayed by those with high $r$ values. Almost all the NMs having $r$ values from 0.88 to 0.90 are in the low latitude region in the Northern hemisphere. However, CAPS, INVK and THUL are at the high latitude in the Northern hemisphere while KGSN and NWRK are at mid latitude in the Southern and Northern hemisphere respectively.

\begin{table}
	\caption{Full names of the place where the neutron monitors for GLE60 are located, their short names and their correlation coefficient ($r$ values) of the PC1}
	\label{table:7}
	\centering
	\begin{tabular}{rlll}
		\hline
	SN. & NM full name & $NM short name$ & $r$ \\ 
		\hline
	1 & APATITY & APTY & 0.88\\
	2 & CALGARY & CALG & 0.89\\
	3 & CAPESSHMIDT & CAPS & 0.88\\
	4 & CLIMAX & CLMX & 0.90\\
	5 & FORTSMITH & FSMT & 0.40\\
	6 & HERMANUS & HRMS & 0.80\\
	7 & INUVIK & INVK & 0.90\\
	8 & IRKUTSK2 & IRKT2 & 0.90\\
	9 & IRKUTSK & IRKT & 0.89\\
	10 & JUNGFRAUJOCH & JUN1 & 0.80\\
	11 & KERGUELEN & KERG & 0.88\\
	12 & KINGSTON & KGSN & 0.90\\
	13 & KIEL & KIEL & 0.94\\
	14 & LOMNICKYSTIT & LMKS & 0.90\\
	15 & MCMURDO & MCMD & 0.48\\
	16 & MAGADAN & MGDN & 0.60\\
	17 & MAWSON & MWSN &  0.66\\
	18 & NAIN & NAIN & 0.78\\
	19 & NORILSK & NRLK & 1.00\\
	20 & NOVOSIBIRSK & NVBK & 0.18\\
	21 & NEWARK & NWRK & 0.40\\
	22 & OULU & OULU & 0.89\\
	23 & SANAE & SNAE & 0.76\\
	24 & SOUTHPOLE & SOPO & 0.78\\
	25 & TERREADELIE & TERA & 0.90\\
	26 & THULE & THUL & 0.89\\
	27 & TIXIEBAY & TXBY & 1.00\\
	28 & YAKUTSK & YKTK & 0.89\\
	
	\hline
\end{tabular}
\end{table}

\subsection{CORRELATION OF THE SECOND PRINCIPAL COMPONENT (PC2) WITH RAW DATA IN GLE 60}

\begin{figure}[htbp]
	\centerline{\includegraphics[width = 0.8\textwidth]{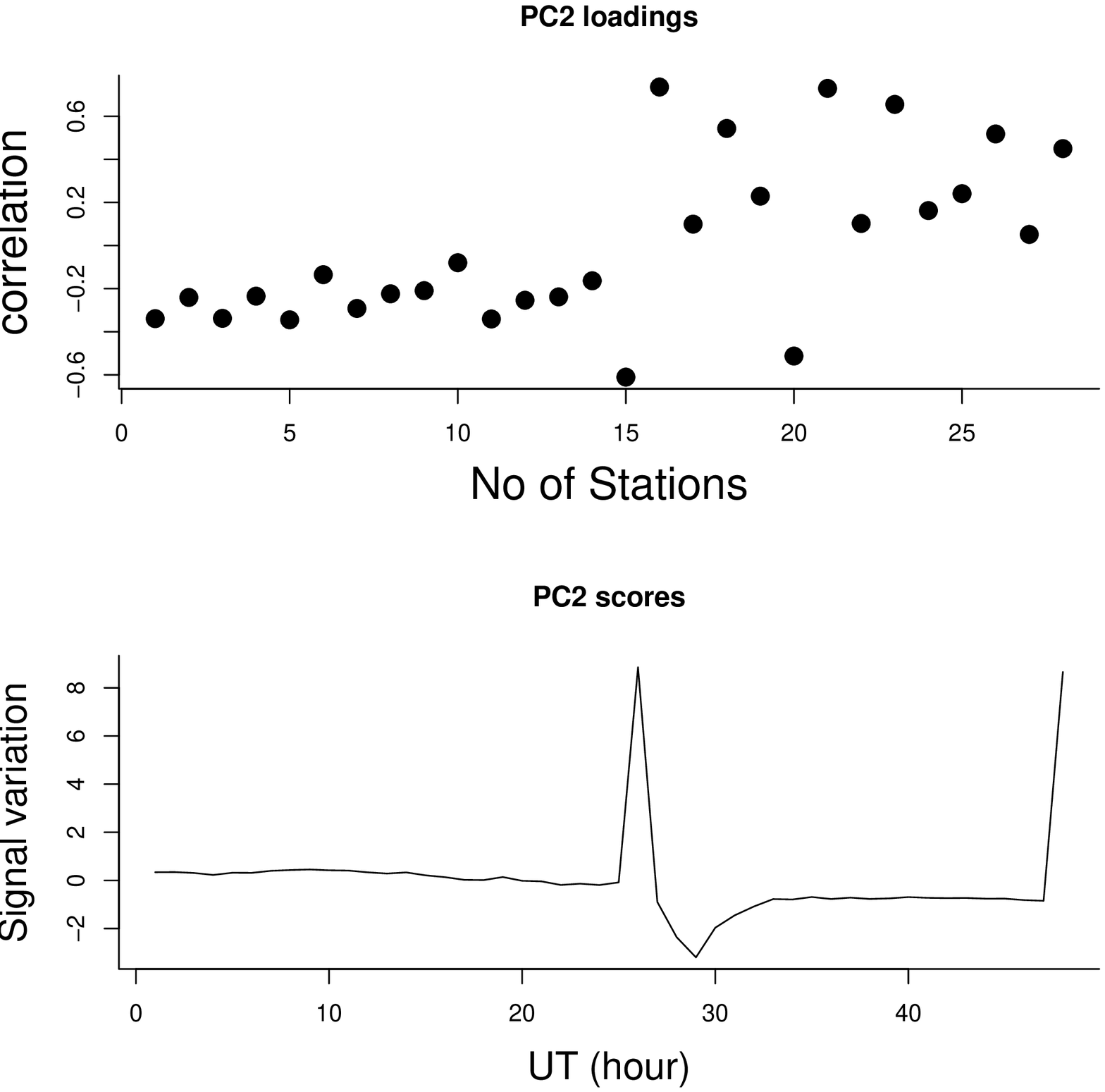}}
	\caption{PC2 for GLE of April 15, 2001 event.}
	\label{Fig. 9}
\end{figure}

PC2 (Figure \ref{Fig. 9}) in this event accounted for 14.26\% of the total intensity variation in the raw data. NMs in PC2 for this event having high $r$ are also the ones that had low $r$ values in PC1. In this PC2, NMs with positive high $r$ values had an instant rise to peak count and their GCR count before the GLE were relatively stable(flat). These NMs and their $r$ are; NAIN (0.46), THUL (0.50), SNAE (0.60), MGDN (0.75), and NWRK (0.75). Those having negative high $r$ did not have the same stable GCR count before and after the GLE even though they had a sharp rise to peak count. These NMs and their $r$ are: NVBK (-0.58) and MCMD (-0.65).

APTY, CAPS and KERG have $r$ = -0.35. Three of them had pre-peak count which possibly could be a sign of a double peak structure that did not show up in hourly plots. YKTK having $r$ = 0.40 also had what looks like double peaks. The difference between it and these three with $r$ = -0.35 is that its second count at the peak is lesser than the first while in those three above, the first count was less than the second.

SOPO and MWSN had $r$ = 0.1. They both have similar profiles characterized by stable (or flat) GCR count before and after the GLE. This means that there was no FD in them before the GLE. They also had an instant rise to peak count. JUN1 which had $r$ = -0.1 had an unstable GCR count before the GLE and after the decay. It showed a large FD and after the recovery phase of the FD, the GLE began with a gradual increment in the CR count for several hours before jumping to a peak count. The following NMs with $r$ = -0.2 have similar CR signal counting structures as JUN1; LMKS, IRKT, and IRK3.

KGSN, INVK and CALG have $r$ = -0.25. They had stable GCR counts before and after the GLE. Thus for several hours before the GLE they did not observe small or large FD. They too had an instant rise to peak count. KIEL with $r$ = 0.3 is the only NM having reversed $r$ value compared to these NMs with $r$ = -0.25. KIEL did not show a stable GCR count both before and after the GLE. It had at least two small FDs just before the GLE.

\begin{table}
	\caption{PCA summary of the April 15, 2001 event (GLE 60)}
	\label{table:8}
	\centering
	\begin{tabular}{rllll}
		\hline
	SN. & Principal Component & Standard deviation & Proportion of Variance & Cumulative Proportion\\ 
		\hline
1 & PC1 &  4.2553           &  0.6467  &               0.6467\\
2 & PC2 &  1.9984           &  0.1426  &               0.7893\\
3 & PC3 &  1.7582           &  0.1104  &               0.8997\\
4 & PC4 &  1.20287          &  0.05167 &               0.95140\\
5 & PC5 &  0.96476          &  0.03324 &               0.98464\\
6 & PC6 &  0.45727          &  0.00747 &               0.99211\\
7 & PC7 &  0.29061          &  0.00302 &               0.99513\\           
8 & PC8 &  0.22915          &  0.00188 &               0.99700\\       
9 & PC9 &  0.1909           & 0.00130  &               0.9983\\
10 & PC10 & 0.11587         & 0.00048  &               0.99878\\            
11 & PC11 & 0.08987         & 0.00029  &               0.99907\\             
12 & PC12 & 0.07865         & 0.00022  &               0.99929\\             
13 & PC13 & 0.06683         & 0.00016  &               0.99945\\              
14 & PC14 & 0.06054         & 0.00013  &               0.99958\\       
15 & PC15 & 0.04928         & 0.00009  &               0.99967\\
16 & PC16 & 0.04596         & 0.00008  &               0.99974\\
17 & PC17 & 0.04212         & 0.00006  &               0.99981\\              
18 & PC18 & 0.03913         & 0.00005  &               0.99986\\
19 & PC19 & 0.03128         & 0.00003  &               0.99990\\
20 & PC20 & 0.02708         & 0.00003  &               0.99992\\
21 & PC21 & 0.02525         & 0.00002  &               0.99995\\
22 & PC22 & 0.02095         & 0.00002  &               0.99996\\
23 & PC23 & 0.01834         & 0.00001  &               0.99997\\
24 & PC24 & 0.01754         & 0.00001  &               0.99999\\
25 & PC25 & 0.01289         & 0.00001  &               0.99999 \\
26 & PC26 & 0.0117          & 0.0000   &               1.0000\\
27 & PC27 & 0.008066        & 0.0000   &               1.000000\\
28 & PC28 & 0.006679        & 0.0000   &               1.000000\\

 \hline
\end{tabular}
\end{table}

\subsection{CORRELATION OF THE FIRST PRINCIPAL COMPONENT (PC1) WITH RAW DATA IN JANUARY 20, 2005 (GLE 69)}
\begin{figure}[htbp]
	\centerline{\includegraphics[width = 0.8\textwidth]{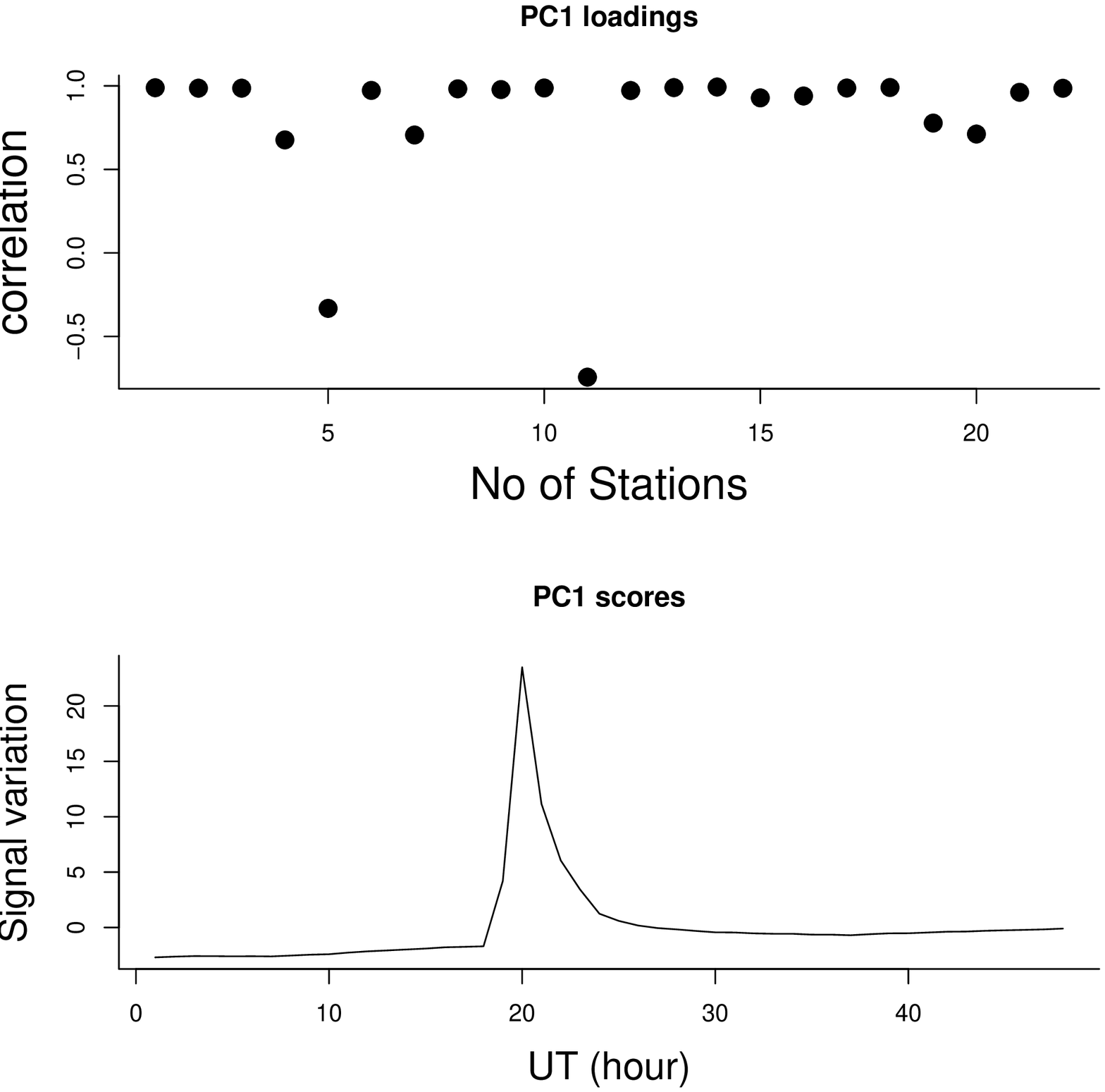}}
	\caption{PC1 for GLE of January 20, 2005 event}
	\label{Fig. 10} 
\end{figure}
Figure \ref{Fig. 10} is the PC1 of the January 20, 2005 GLE. Tables \ref{table:9} and \ref{table:10} show the quantitative results of the GLE. The PCA summary for this event in Table \ref{table:10} shows that the proportion of variance for PC1 is 0.8181, implying that PC1 accounts for 81.8\% of the total intensity variation in the raw data. In comparison with the event of 29 September, 1989, the presented results of PCA confirm that the GLE of 20 January 2005 is smaller \citep[see][]{Mc:2014}.  

Fourteen out of twenty-two NMs considered for PCA have $r$ = 1.00 in the PC1 and their peak occurred at 07:00 UT. This is about 63\%. The NMs which included APTY, CALG, CAPS, INVK, KERG, KGSN,KIEL, MGDN, MWSN, NAIN, OULU, SNAE, THUL, YKTK, and NWRK all had similar decay phase up to the day after the event. In other words their recovery phase were also similar. All the profiles show quasi-exponential decay. Though there are noticeable differences in their rise to peak count, their decay count are so similar that they are not distinguishable. Seven of these NMs are at the high latitude in the Northern hemisphere, four in the mid latitude in the Northern hemisphere, two at the mid and high latitudes in the Southern hemisphere.

Most of the NMs whose $r$ values are equal to 1.00 had initial rise in CR count that are relatively small. The increase is much less in INVK, CAPS, NWRK,THUL, MGDN than in others.   Their profile also show that the GLE began with an FD as a precursor. The bigger the pre increase, in the CR count, the larger the FD. Those that revealed a small FD beside the large FD (example MGDN, SOPO and THUL) did not show the initial rise in the CR count instead, they had instant jump to the peak count.

NRLK and NVBK had $r$ = 0.95 and the profiles from both stations peak at 07:00 UT. Their decay phase are similar and slightly different from those with $r$ = 1.00 above. They also had initial increase in CR counts after the recovery phase of the FD before an instant jump to the peak. The steepness of these precursory counts was more in NVBK than in NRLK. NRLK equally had a pre peak count which is not detected in NVBK.

IRK2  had $r$ = 0.70. It never recovered like the previous ones considered before there was rise in CR count the same day. It had its peak at 06:00 UT. The decay in CLMX with $r$ = 0.65 was below the initial background GCR counts. This made it quite different from others. FSMT and MCMD  unlike any other NM recorded negative correlation factor with $r$ = -0.40 and -0.70  respectively. They recorded FD when others were observing enhancement of CRs. MCMD had greater value of negative correlation because after the maximum depression or drop in the CR counts , it tried to recover but the count dropped again to almost the maximum depression. From this later point, it registered an enhancement. MCMD therefore recorded two FDs and first one is the larger FD.  

Most of the stations having $r\geq 0.95$ are at high latitudes ($60.10^{\circ}$-$76.50^{\circ}$) North of the hemisphere except NVBK (latitude $54.80^{\circ}$). All NMs that had two steps rise to peak also had $r\geq 0.95$. The two among them that have $r$ = 0.95 (NRLK and NVBK) also show identical CR count signal. MWSN  at high latitude ($-67.60^{\circ}$) in the Southern hemisphere has $r$ = 1.00. SOPO at latitude $-90.0^{\circ}$ and TERA at latitude $-66.70^{\circ}$ also have $r$ = 0.80 and 0.70, respectively. They both have instant rise to peak count that occurred at 06:00 UT. SOPO did not show an FD prior to the GLE. Stations that show instant rise to the CR peak count also have $r\leq 0.80$ in GLE 69.

\begin{table}
	\caption{Full names of the place where the neutron monitors for GLE69 are located, their short names and their correlation coefficient ($r$ values) of the PC1}
	\label{table:9}
	\centering
	\begin{tabular}{rlll}
		\hline
	SN. & NM full name & $NM short name$ & $r$ \\ 
		\hline		
		
	1 & APATITY & APTY & 1.00\\
	2 & CALGARY & CALG & 1.00\\ 
	3 & CAPESSHMIDT & CAPS & 1.00\\
	4 & CLIMAX & CLMX & 0.65\\
	5 & FORTSMITH & FSMT & -0.40\\
	6 & INUVIK & INVK & 1.00\\
	7 & IRKUTSK2 & IRK2 & 0.70\\ 
	8 & KERGUELEN & KERG & 1.00\\
	9 & KINGSTON & KGSN & 1.00\\
	10 & KIEL & KIEL & 1.00\\
	11 & MCMURDO & MCMD & -0.70\\
	12 & MAGADAN & MGDN & 1.00\\                        
	13 & MAWSON & MWSN & 1.00\\                                              
	14 & NAIN & NAIN & 1.00\\                          
	15 & NORILSK & NRLK & 0.95\\
	16 & NOVOSIBIRSK & NVBK & 0.95\\                                  
	17 & NEWARK & NWRK & 1.00\\                                  
	18 & OULU & OULU & 1.00\\                                                                                         
	19 & SOUTHPOLE & SOPO & 0.80\\
	20 & TERREADELIE & TERA & 0.70\\
	21 & THULE & THUL & 1.00\\                                                      
	22 & YAKUTSK & YKTK & 1.00\\

	\hline
\end{tabular}
\end{table}

\subsection{CORRELATION OF THE SECOND PRINCIPAL COMPONENT (PC2) WITH RAW DATA IN GLE 69}

\begin{figure}[htbp]
	\centerline{\includegraphics[width = 0.8\textwidth]{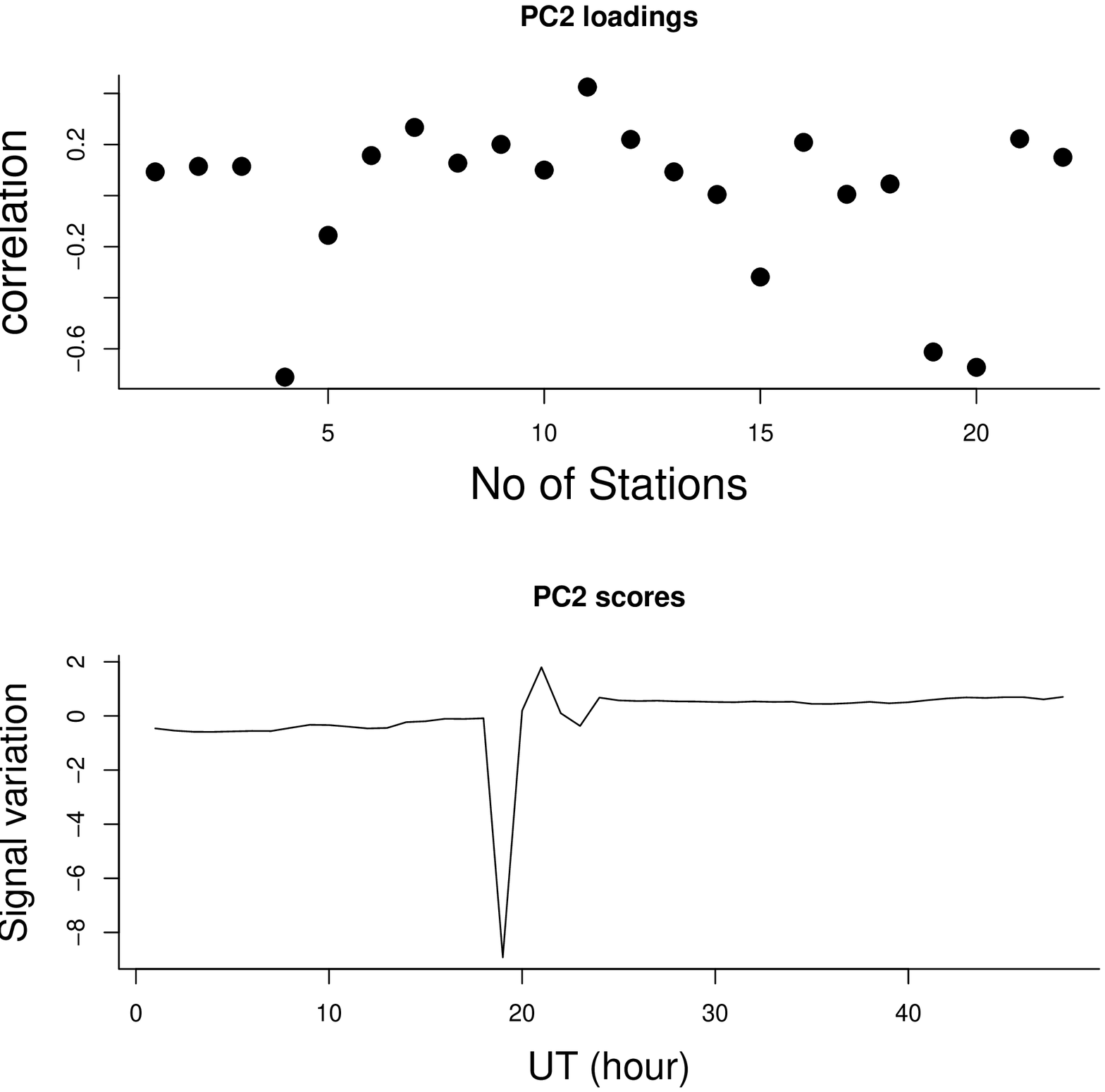}}
	\caption{PC2 for GLE of January 20, 2005 event.}
	\label{Fig. 11}
\end{figure}

PC2 (Figure \ref{Fig. 11}) in this event accounts for 9.14\% of the total intensity variation in the raw data. It has always been the case that close values of has $r$ reflect greater resemblance of the  structure of the CR count. In this PC2, those with has $r$ = 0.1 such as APTY, CALG, CAPS, KERG, KIEL, MWSN and YKTK had close CR count signal. While CAPS and INVK rose to peak counts instantly, the rest had initial small increase in CR count before jumping to the peak count. FSMT is the only NM with close negative has $r$ (-0.2) as these NMs. It had a complete large FD at the period others were having GLE. Its FD is a reverse of the GLE in terms of its CR counts. First, it dropped to a certain level before its maximum decent in the CR count just as majority of these NMs first had an initial increase in CR count before the peak count. NAIN (whose has $r$ = 0.00) and other  NMs with has $r$ = 0.2 (MGDN, NVBK and THUL) had similar CR count signal structure as those with has $r$ = 0.1. 

While MCMD has $r$ = 0.40, NRLK has $r$ = -0.40. NRLK is characterized by a large pre-peak count and later a peak count. At the same time, MCMD had a large FD. Even when it tried to recover, it dropped again to almost the level before the onset recovery. After its recovery, the observed GCR count was higher than what it was before the GLE event. The decay of NRLK also established a new level of GCR count that is higher than what it was before the GLE.

SOPO, TERA and CLMX have $r$ = -0.65, -0.68 and -0.70 respectively. There were no other NMs with high negative $r$ value that could be compared with these values.

\begin{table}
	\caption{PCA summary of the January 20, 2005 event}
	\label{table:10}
	\centering
	\begin{tabular}{rllll}
		\hline
	SN. & Principal Component & Standard deviation & Proportion of Variance & Cumulative Proportion\\ 
		\hline

1 & PC1 &  4.2424     &             0.81810        &        0.8181\\
2 & PC2 &  1.41786    &             0.09138        &        0.90947\\
3 & PC3 &  0.97242    &             0.04298        &        0.95245\\
4 & PC4 &  0.80133    &             0.02919        &        0.98164\\
5 & PC5 &  0.52269    &             0.01242        &        0.99405\\
6 & PC6 &  0.25495    &             0.00295        &        0.99701\\                                  
7 & PC7 &  0.15675    &             0.00112        &        0.99813\\         
8 & PC8 &  0.12213    &             0.00068        &        0.99880\\                                                       
9 & PC9 &  0.10396    &             0.00049        &        0.99929\\                  
10 & PC10 &  0.07308  &             0.00024        &        0.99954\\                   
11 & PC11 &  0.06076  &             0.00017        &        0.99971\\
12 & PC12 &  0.04694  &             0.00010        &        0.99981\\                                   
13 & PC13 &  0.03855  &             0.00007        &        0.99987\\                    
14 & PC14 &  0.03397  &             0.00005        &        0.99993\\                                   
15 & PC15 &  0.02303  &             0.00002        &        0.99995\\                 
16 & PC16 &  0.02006  &             0.00002        &        0.99997\\                                    
17 & PC17 &  0.01595  &             0.00001        &        0.99998\\                                  
18 & PC18 &  0.01368  &             0.00001        &        0.99999\\                                    
19 & PC19 &  0.01212  &             0.00001        &        0.99999\\                  
20 & PC20 &  0.00917  &             0.00000        &        1.00000\\                                     
21 & PC21 &  0.006379 &             0.00000        &        1.000000\\                                    
22 & PC22 &  5.17e-18 &             0.00000        &        1.000000\\                                               

\hline
\end{tabular}
\end{table}

\section{SUMMARY AND CONCLUSION}
\label{sect:conclusion}
We have applied PCA to the ground level enhancement of CR events of May 7, 1978; September 29, 1989; October 22, 1989; April 15 2001 and January 20 2005. In this study, a combination of PC1 and PC2 sufficiently accounted for the information in the raw data in each of the event. 

Our application of PCA to GLE studies yielded the following information:
\begin{itemize}
\item The PCA is well suited for the investigation of GLEs across multiple NM stations. It can, for example, clearly distinguish between strong/globally simultaneous GLEs and weak/non-simultaneous ones.
\item The time profile and structure of the CR count signal determine which of them will have the same or close values of $r$. In all the events, almost all the NMs with the same $r$ values have the same time of peak. Having the same time of peak may not lead to equivalent $r$ value but having the same $r$ value in most cases resulted in the same time of peak. 
\item NMs with identical signal structures of CR counts are assigned the same $r$ value.
\item Most of the NMs that had the same $r$ value tend to occupy close latitude range.
\item NMs that had FD at the time others observed GLE in an event can be distinguished from others as they usually have negative correlation factor as against others that have positive correlation factor. 
\item NMs with close $r$ values equally have very similar CR profiles.
\item The features that are common to most or all the NMs in an event is what makes $r$ to be higher in all the NMs. Such feature could be that majority of the NMs had instant rise to peak count or that they had the same decay pattern or that they all had the same pre-peak count or a combination of two or more of these features.

\item  Analysing PC2 requires making comparison between NMs having the same or close positive $r$ with NMs having the same or close values of negative $r$. It usually yields information on the differences between NMs having opposite but the same or close values of $r$.
The presented result of PCA on GLE is preliminary. It has the potential of stimulating a more detailed future work. 

\item One of the major flaws of the PCA tool in the CR investigation is its inability to account for CR diurnal anisotropies in raw CR data. CR diurnal anisotropy has significant influence on the time-intensity profiles/structure of CR signal. The results presented here should, therefore, be viewed with caution. In a future work, efforts will be made to remove the contributions of CR anisotropy before passing the data to the PCA algorithm. 
Our investigation has shown that PCA is an essential technique for analysing multivariate GLE data. GLE research that involve several NM data should even begin with PCA because a lot of shared information by the NMs are promptly thrown up by the PCA.

\end{itemize}

\section {Acknowledgments}
We are indebted to the Pushkov Institute of Terrestrial Magnetism, Ionosphere, and Radio Wave Propagation, Russian Academy of Sciences (IZMIRAN) who host the website \url{http://cro.izmiran.ru/common/links.htm}. The raw data used in this investigation are from their data bank. This work was sponsored by the Nigeria Tertiary Education Trust fund (TETFUND) implemented by Federal University of Technology Owerri, Imo State Nigeria. We are indebted to the anonymous referee. His/her inputs greatly changed the manuscript.


\bibliography{reference}

\end{document}